\documentclass[aip,reprint]{revtex4-1}

\usepackage{lipsum}
\usepackage{amsmath}
\usepackage{amssymb}
\usepackage{graphicx}
\usepackage{gensymb}
\usepackage{tabularx}
\usepackage{float}
\restylefloat{table}
\DeclareMathOperator\erf{erf}
\newcolumntype{Y}{>{\centering\arraybackslash}X}
\begin{document}
	\title{Synthetic Nuclear Diagnostics for Inferring Plasma Properties of Inertial Confinement Fusion Implosions}
	\author{A. J. Crilly}
	\author{B. D. Appelbe}
	\author{K. McGlinchey}
	\author{C. A. Walsh}
	\author{J. K. Tong}
	\author{A. B. Boxall}
	\author{J. P. Chittenden}
	\affiliation{1. Centre for Inertial Fusion Studies, The Blackett Laboratory, Imperial College, London SW7 2AZ, United Kingdom}

	\begin{abstract}
		A suite of synthetic nuclear diagnostics has been developed to post-process radiation hydrodynamics simulations performed with the code Chimera. These provide experimental observables based on simulated capsule properties and are used to assess alternative experimental and data analysis techniques. These diagnostics include neutron spectroscopy, primary and scattered neutron imaging, neutron activation, $\gamma$-ray time histories and carbon $\gamma$-ray imaging. Novel features of the neutron spectrum have been analysed to infer plasma parameters. The nT and nD backscatter edges have been shown to provide a shell velocity measurement. Areal density asymmetries created by low mode perturbations have been inferred from the slope of the downscatter spectrum down to 10 MeV. Neutron activation diagnostics showed significant aliasing of high mode areal density asymmetries when observing a capsule implosion with 3D multimode perturbations applied. Carbon $\gamma$-ray imaging could be used to image the ablator at high convergence ratio. Time histories of both the fusion and carbon $\gamma$ signals showed a greater time difference between peak intensities for the perturbed case when compared to a symmetric simulation.
	\end{abstract}
	
	\maketitle
	
	\section{Introduction}
	
	Ignition of Inertial Confinement Fusion (ICF) capsules favours a spherically symmetric implosion and stagnation phase \cite{lindl1992}. Perturbations seeded by X-ray drive asymmetry, surface roughness and engineering features cause loss of this symmetry. Observing the effects of these perturbations is key in identifying the principal cause of failed ignition. Simulations with detailed information of the hydrodynamic quantities of the stagnated capsule plasma can be used to investigate the effect of perturbations on performance. Synthetic diagnostics can be constructed to produce equivalent measurements of observable experimental signatures from the simulated implosion. This allows for a more direct comparison between simulation and experiment. Previous literature in this area investigated X-ray and neutron diagnostic signatures from 3D simulations of ICF implosions \cite{Taylor2013,Chittenden2016,Spears2015,Weber2015}. Fluid velocity introduced by drive asymmetries was found to increase the inferred ion temperature and cause variation in this temperature along different lines of sight. This variation is also observed within experiments \cite{GatuJohnson2016,Spears2015}. \\
	
	Synthetic diagnostics also provide an avenue to investigate perturbation effects in isolation and develop new analysis techniques for use on experimental data. In this work, 3D radiation hydrodynamics simulations will be post-processed to obtain current and novel experimental observables. These diagnostic signatures will be used to infer plasma properties which are then compared to those within the hydrodynamic simulation.\\
	
	Indirect drive experiments of layered deuterium-tritium capsule implosions at the National Ignition Facility (NIF) primarily observe neutrons, X-rays and $\gamma$-rays. The spectrum of neutron energies contains information about both the plasma conditions where the fusion reactions are occurring \cite{Appelbe2011,Munro2016} as well as nuclear interactions taking place on the neutron's flight to the detector \cite{Johnson2012,Frenje2010}. Inhomogeneous and time-varying plasma conditions will affect the rate and type of interactions and hence the resulting spectra. Spectroscopy is performed experimentally by the Neutron Time of Flight detectors (nToFs)\cite{Glebov2010,Glebov2012} and the Magnetic Recoil Spectrometer (MRS)\cite{Casey2013}. Measurements by these diagnostics are used to infer ion temperatures, areal densities and fluid motion \cite{Munro2016,Munro2017,Chittenden2016,Johnson2012,GatuJohnson2016,Hatarik2015}. These are generally calculated from primary and low scattering angle features of the spectra. Imaging can be used to find spatial information about the emitting plasma. Primary neutrons are produced within the hotspot and the majority of scattered neutrons within the cold dense DT fuel, which will be referred to as the shell. The Neutron Imaging System (NIS)\cite{Merrill2012,Grim2013} is used to infer hotspot and cold shell shape \cite{Grim2013,Volegov2015} and the flange-mounted Neutron Activation Diagnostic system (FNADs) infers the areal density variations over 4$\pi$ of solid angle via measurement of the attenuated primary neutron flux \cite{Bleuel2012,Yeamans2012,Yeamans2017}. Gamma rays are produced in fusion reactions and when neutrons inelastically scatter from carbon within the ablator. The $\gamma$-Ray History detector (GRH) measures the time history of $\gamma$-ray production which is used to measure the time of peak neutron production (referred to as bang time) and carbon-based ablator areal density \cite{Cerjan2015,Hoffman2013}. Images of the carbon $\gamma$-rays could show the spatial distribution of ablator, and when combined with neutron images this creates a complete picture of hotspot, shell and ablator structures. \\
	
	In this paper, we will discuss the development of synthetic diagnostics for neutrons and $\gamma$-rays. Various numerical methods exist for the transport of neutrons through reactive media. Typically stochastic Monte Carlo methods are employed for neutron transport in ICF plasmas, for example in the MCNP code \cite{MCNP}. Alternatively there are deterministic methods which aim to numerically solve a discretised transport equation for the particle flux throughout the phase space of the problem. In this work, results produced by deterministic methods will be investigated. These include a 3D ray tracer limited to single scattering and a more accurate 1D discrete ordinates method used to estimate the errors of the ray tracer. \\
	
	The neutron transport models discussed within this work are used to post-process radiation hydrodynamics simulations performed by Chimera \cite{Chittenden2016}. Chimera is a 3D Eulerian radiation magnetohydrodynamics code with non-diffusive multigroup radiation transport \cite{McGlinchey2017} using atomic data provided by SpK \cite{Niasse2012}, Spitzer-H\"{a}rm thermal transport \cite{Spitzer1953}, extended MHD capabilities \cite{Walsh2017} and equation of state tables from FEoS \cite{FEOS1,FEOS2}. All simulations considered in this investigation are based on the high performing High-Foot shot N130927 \cite{Park2014,Hurricane2014}. 3D simulations with long wavelength perturbations were constructed from a library of 1D simulations via a Legendre polynomial decomposition of the radiation field, as in Chittenden et al.\cite{Chittenden2016}. This library of 19 1D simulations was performed with radiation temperature varied in the range $\pm$ 4.5 \%  \cite{Chittenden2016}. Short wavelength perturbation simulations were created by initialising a nominal 1D simulation in 3D at peak implosion velocity and applying velocity perturbations consistent with the Rayleigh-Taylor instability \cite{Layzer1955,Taylor2013b,Taylor2013}. The 3D simulations presented here are calculated on regular Cartesian grids with cellsizes of either 1 or 2 $\mu$m and a hydrodynamic time step of order 1 ps.\\
	
	A detailed description of the deterministic code developed to solve the transport of neutrons in 3D is given in section II. Section III presents work on neutron spectroscopic signals; analysis of the backscatter edges and the downscatter spectra down to 10 MeV produced in perturbed simulations are discussed. Section IV presents work on synthetic neutron imaging and image analysis techniques. Finally, $\gamma$-ray images and time histories produced by inelastic neutron scattering from carbon are investigated in section V. The conclusions of this work are discussed in section VI. Description of the 1D code used to benchmark the 3D code and the backscatter edge fitting models are given in the appendices. 
	
	\section{Neutron Transport}
	
	The behaviour of neutrons produced by fusion reactions is described by a transport equation. This characterises the spatial variation, time evolution and energetic distribution of a population of neutrons within a medium with which the neutrons can interact. In capsule implosions, fusion reactions in the hotspot act as the external neutron source. For a deuterium-tritium (DT) capsule, the reactions are between DT, DD and TT, where DT has the largest reactivity and produces a 14 MeV neutron. Interactions of neutrons with ions in the environment can alter the number, energy and direction of the neutrons. The fusion reactions and nuclear interactions relevant to ICF plasmas which have significant cross section include the primary fusion reactions (DT,DD,TT), deuteron break-up (D(n,2n)), elastic and inelastic neutron scattering from D, T, C and H. Total and differential cross sections from the ENDF/B-VII \cite{ENDF} (for nH, nD and nT scattering) and CENDL-3.1 \cite{CENDL} (for D(n,2n) and nC scattering) evaluated nuclear libraries were used in this work. It should be noted that the secondary and reaction-in-flight (RIFs) neutron sources will not be considered. \\
	
	A 14 MeV DT fusion neutron travels at $\sim$ 50 $\mu$m/ps. This is considerably faster than the bulk fluid velocities during neutron production which are $\sim$ 0.1 $\mu$m/ps. Hence the neutronic behaviour can be assumed to be time independent within a static background for each hydrodynamic time step. \\
	
	\begin{figure}[htp]
		\centering
		\includegraphics*[width=0.495\textwidth]{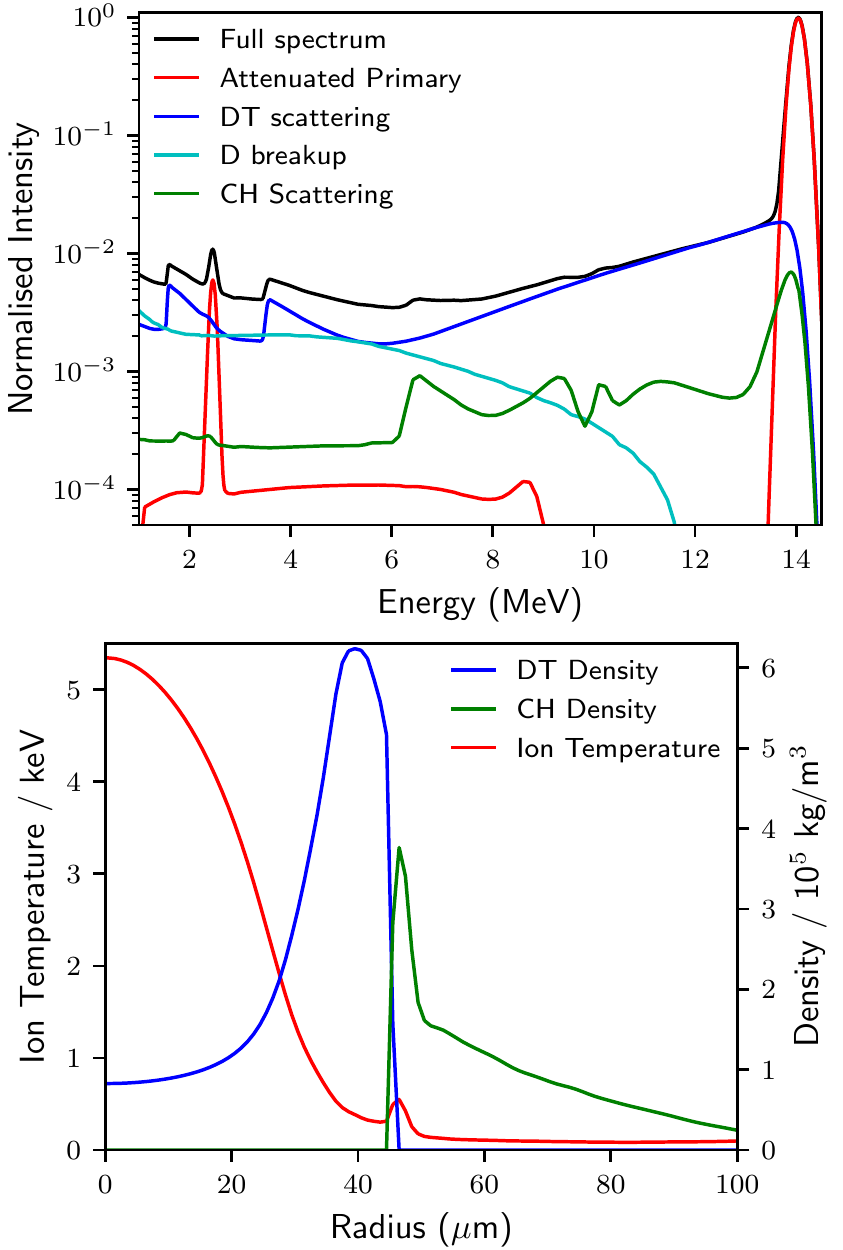}
		\caption{Top: Neutron spectrum created by post-processing a 1D Chimera simulation of the High-Foot shot N130927 using Minotaur. The contributions to full spectrum, black line, from various nuclear interactions are shown. Multiple scattering events are considered in this calculation. Bottom: The simulated density and temperature radial profiles at bang time for High-Foot shot N130927 \cite{Chittenden2016}. A neutron yield of 7.85 $\times$ 10$^{15}$ was obtained without alpha-heating effects.}
		\label{fig:minotaur}
	\end{figure}

	An example of the full 1D neutron spectrum is given in figure \ref{fig:minotaur}, this was produced using the 1D discrete ordinates code Minotaur, which is described in appendix A. The neutron spectrum shows several features of interest; the DT and DD peaks occurring at $\sim$ 14 and 2.5 MeV respectively, the backscatter edges either side of the DD peak and the spectrum down to 10 MeV which is dominated by scattering from D and T. The 1D neutron transport calculations give knowledge of the full detail of the spectrum for an unperturbed implosion. This provides a reference when evaluating, with a more approximate method, the effect of 3D asymmetries on spectral features.\\

	In order to efficiently produce synthetic diagnostics in 3D, an inverse ray trace scheme was devised. As only diagnostic signatures are of interest, only the neutrons which will arrive at the detector are tracked. This greatly reduces the size of the calculation making it tractable in 3D. Previously, this has been used for images and yield measurements for primary and 10-12 MeV neutrons \cite{Chittenden2016}. This method has since been extended to lower neutron energies for spectral, imaging and yield diagnostics. Using the method of characteristics, the time independent form of the transport equation can be rewritten in a line integral form \cite{Bell_Glasstone_1970}:
	\begin{align}\label{methodofcharacteristics}
	\psi(\vec{r},\hat{\Omega},E) =& \int_0^{\infty}  \exp\left[-\sigma(E)\int_0^{s'} ds''n(\vec{r}-s''\hat{\Omega})\right]\\&S(\vec{r}-s'\hat{\Omega},\hat{\Omega},E) \ ds' \nonumber
	\end{align}
	Where $\psi$ is the angular neutron flux, $E$ and $\hat{\Omega}$ are the energy and direction of motion of the neutrons and $n$ is the number density of the interacting species with total cross section denoted by $\sigma$. $S$ represents both the external and nuclear interaction source terms. Solving along the detector line of sight, $(\vec{r},\hat{\Omega}) = (\vec{r}_{det},\hat{\Omega}_{det})$, now constitutes a single line integral if $S$ is known. In practice this integral tracks neutron paths from the detector plane back through the simulation grid. The source and degree of attenuation is calculated for each grid cell intersected. For a primary source, the source term is simply the reaction rate in that grid cell.\\
	
\begin{figure}[htp]
		\centering
		\includegraphics*[width=0.485\textwidth]{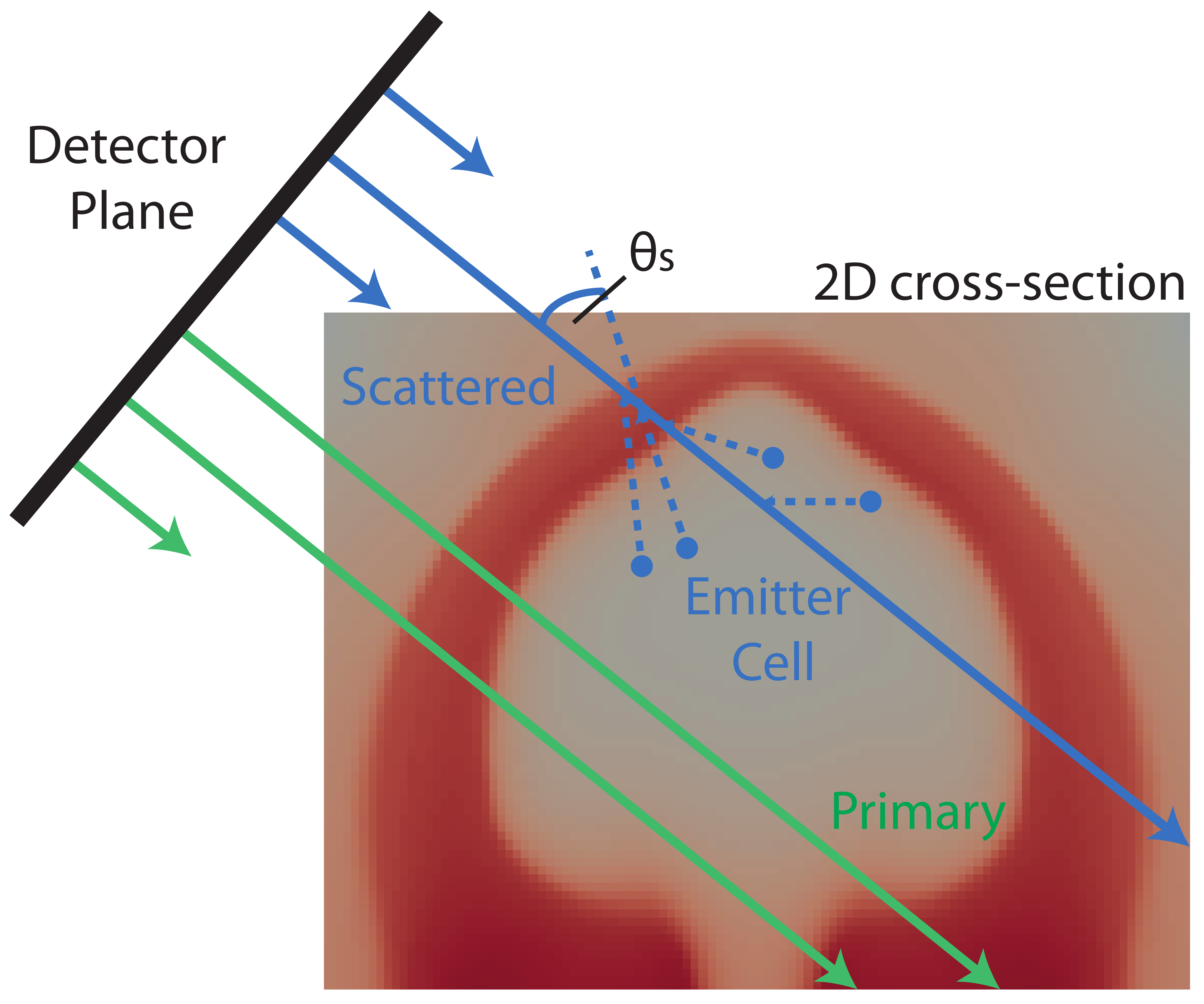}
		\caption{Schematic of the inverse ray trace method. In green, primary neutrons are tracked by straight rays traced back from the detector plane through the simulation grid. In blue, scattered neutrons are found by a combination of many traces. Neutrons are traced from the emitter cells to every scattering cell along the detector line of sight. The range of allowed scattering angles, $\theta_s$, is set by the energy gate considered.}
		\label{fig:raytrace}
	\end{figure}

	Typically $\sim$20\% of neutrons produced in the D(T,n) reaction undergo scattering events. Hence scattering accounts for a significant portion of the neutronic behaviour. Due to its high areal density, the majority of scattering events occur within the shell. Generally these scattering events will be elastic collisions with relatively slow moving ions. For an elastic collision with a stationary ion ($E_{\mbox{n}} \gg$ $E_{\mbox{ion}}$) of mass $A m_n$, the incoming neutron energy, $E'$, and outgoing neutron energy, $E$, can be directly related to the scattering cosine, $\mu \equiv \cos(\theta_s)$:
	\begin{equation}\label{elastic}
	\frac{E}{E'} = \frac{\left(\mu+\sqrt{\mu^2+A^2-1}\right)^2}{\left(A+1 \right)^2}
	\end{equation}
	This relationship between scattered neutron energy and geometry is particularly important when looking at perturbed implosions. Inclusion of ion velocity due to fluid motion and inelastic collisions \cite{Williams1971} are simple extensions to this relation. The inverse ray trace method is outlined diagrammatically in figure \ref{fig:raytrace}. A scattering source is handled via an additional set of traces from the intersected cell to all emitting cells above an emission power threshold. This is a considerably larger calculation than for the primary source. Since each ray from an emitter to the scattering cell is independent, this calculation can be fully parallelised. Generally only a limited scattered energy range is of interest, allowing the calculation size to be reduced by considering only a subset of rays for which the primary neutrons downscatter into the accepted range. \\
	
	A number of approximations are made to improve the efficiency of the calculation. The Brysk form with relativistic corrections\cite{Brysk1973,Ballabio1998} is used for the neutron birth spectrum within each grid cell's rest frame. Transforming to the lab frame results in a loss of the Gaussian form of Brysk. Noting that the fluid velocity is significantly lower than the neutron velocity, the Gaussian form can be recovered via a first order binomial expansion in the ratio of fluid to neutron velocity. The final primary spectrum constitutes a sum of these approximated Brysk spectra. The total primary spectral peak can be described through its cumulants: mean, variance, skew and kurtosis\cite{Munro2016}. Although the individual spectra reaching the synthetic detector possess no higher cumulants than variance, their sum will. Therefore, the moment analysis outlined by Munro \cite{Munro2016,Munro2017} is still possible with some error introduced. No cumulants higher than variance will be considered in this paper. By considering Gaussians, only three variables are required to be transported along the rays: amplitude, mean and variance. This is computationally efficient compared to a multigroup treatment. Additionally, the Gaussian form can be exploited when handling attenuation. Performing a piecewise linear fit to the total cross section centred on the spectral mean simplifies the calculation of the neutron attenuation to a mean shift \cite{Munro2016} and amplitude reduction. For scattering, the energy dependence of the differential and absolute cross section is assumed constant about the spectral mean. This allows the scattered spectrum mean and variance on the ray to be simply related to the source mean and variance via a multiplicative factor given by equation \ref{elastic}.
	
	Currently, only singly-scattered and un-scattered neutrons are included for the inverse ray trace method. This approximation introduces significant error at lower neutron energies as the degree of multiple scattering increases. However, features at higher neutron energies and distinct single scattering phenomena such as the backscatter edges can be analysed within this approximation with appropriate background subtraction. Minotaur was used to benchmark the more approximate 3D ray trace calculations by ensuring agreement on a symmetric implosion. The background signal created by multiple and ablator scattering and break-up reactions was quantified. For expedience these backgrounds are omitted in 3D calculations.\\
	
	Properties inferred from neutronic measurements will be ``neutron-averaged" quantities. These quantities are what a neutron sees on average on its path from production to detection and will be denoted with angle brackets, $\langle\rangle$. These averages are calculated with a weighting function given by the number of neutrons undergoing the process which creates the observed signal. As an example, the neutron-averaged areal density, $\langle \rho$R$\rangle$ is calculated by an integration over space and time of the product of the areal density from a point to the detector and the production rate of neutrons at that point divided by the neutron yield.\\
	
	\section{Neutron Spectroscopy}

	Analysis of the primary spectrum is well described through a moments approach \cite{Munro2016}, but other spectral features provide additional information on plasma properties. In this paper we will consider the 10 - 12 MeV region and the backscatter edges in detail. These features are seen within experimentally measured spectra \cite{Glebov2012}.
	
	\subsection{Analysis of Backscatter Feature}
	
	When neutrons undergo a 180$\degree$ elastic scattering event, their energy is reduced by a factor dependent on the mass ratio of neutron to scatterer. For nD and nT scattering these are approximately 9 and 4 respectively, which can be seen from equation \ref{elastic}. For a DT fusion neutron source, the nD and nT edges will occur at 1.57 and 3.53 MeV respectively.\\
	
	With a mono-energetic source these backscatters will show up as a sharp edge in the spectrum. A distribution of primary energies due to hotspot temperature\cite{Brysk1973,Ballabio1998} or velocity variance\cite{Murphy2014} will smooth this edge. If the scattering medium has significant fluid velocity it will influence the scattering kinematics for neutrons, causing a shift in the position of the backscatter edge; for the case of collinear collisions:
	\begin{equation}\label{eqnalpha}
		\left(\frac{E}{E'}\right)_{bs} = \left(\frac{A-1+2Av_f/v'_n}{A+1}\right)^2
	\end{equation}
	Where $v_f$ is the fluid velocity of the scatterer and $v'_n$ is the incident neutron velocity, the other variables are defined equation in \ref{elastic}. Since the majority of scattering occurs within the shell, the velocity inferred from the shift in the edge will be strongly weighted towards the fluid velocity in the shell. By using equation \ref{eqnalpha}, assuming a Gaussian birth spectrum and neglecting the angular dependence of the differential cross section, simplified models to fit the backscatter edge spectral shape and infer values of hotspot velocity, temperature and shell velocity are derived and tested in Appendix B. Two models are presented: model \ref{eqnedge1} which includes the effect of hotspot temperature and velocity as well as a single average shell velocity, and model \ref{eqnedge2} which extends the previous model to include a Gaussian distribution of shell velocities. In this section we will apply these models to results of a Chimera simulation. \\
	
\begin{figure}[htp]
	\centering
	\includegraphics*[width=0.495\textwidth]{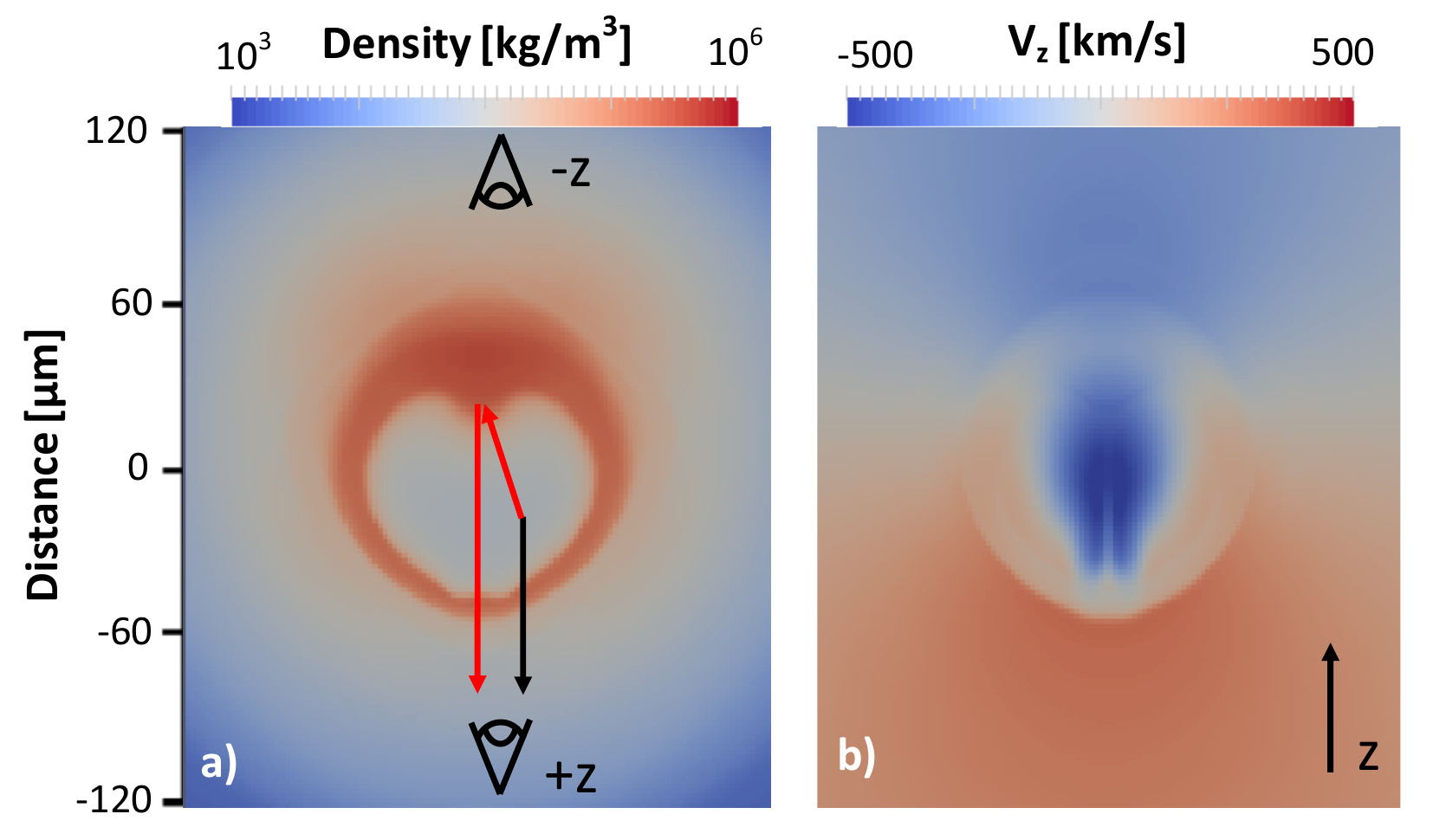}
	\caption{P1 simulation at bang time. \textbf{a)} Density cross section showing hotspot and perturbed dense shell. For a single fluid element, the paths of primary neutrons (black) and backscatter neutrons (red) are shown. Primaries are seen parallel to the fluid velocity, and backscattered neutrons are emitted anti-parallel to the hotspot flow. Thus they will have oppositely directed Doppler shifts. An additional up shift in energy will be caused by the fluid velocity of the shell which is parallel to the detector direction. Detector lines of sight are shown and labelled. \textbf{b)} $z$-component of fluid velocity within the x-z plane showing large bulk fluid velocity along -$z$.}
	\label{fig:P1_diagram}
	\end{figure}

	3D perturbations result in residual kinetic energy in the hotspot and shell. Bulk fluid velocity will alter an observed primary spectrum via a Doppler shift. The scattering medium will also see a shifted birth spectrum depending on its relative velocity. Large fluid velocities in this medium will affect the energy of the backscattering neutrons. A strong feature in the scattered spectrum could be used to probe the birth spectra seen by the scattering sites and the fluid velocity of the scattering medium. The nT backscatter edge is a good candidate as it is a single-scattering event phenomenon, and so can be investigated using our inverse ray trace code. As an example, the time integrated spectrum from a capsule implosion with a P1 asymmetry will be considered. The magnitude of the drive asymmetry, 3\% P1/P0, used in the Chimera simulation caused a 60\% reduction in neutron yield compared to the symmetric case. A 2D slice in the x-z plane of the simulation at bang time is shown in figure \ref{fig:P1_diagram}. This presents a scenario where, when viewed along the +z axis (denoted as $+$z line of sight), the detected primary neutrons are shifted up in energy, but the neutrons travelling towards the back of the capsule are down shifted. The backscattered neutrons will then receive a positive energy shift from the fluid velocity of the shell. Hence the shifts from the hotspot and from the shell are oppositely directed. When viewed along the opposite direction ($-$z line of sight) both the shifts act to increase the backscattering neutron energy, albeit with a different shell velocity magnitude. Combining spectra from antipodal lines of sight will allow the calculation of separate hotspot and shell velocities. The primary spectra measured constrain the spectrum of neutrons arriving at the backscattering sites. As a result no additional assumptions need to be made about the birth spectrum used to fit the edges. \\
	
\begin{figure}[htp]
	\centering
	\includegraphics*[width=0.495\textwidth]{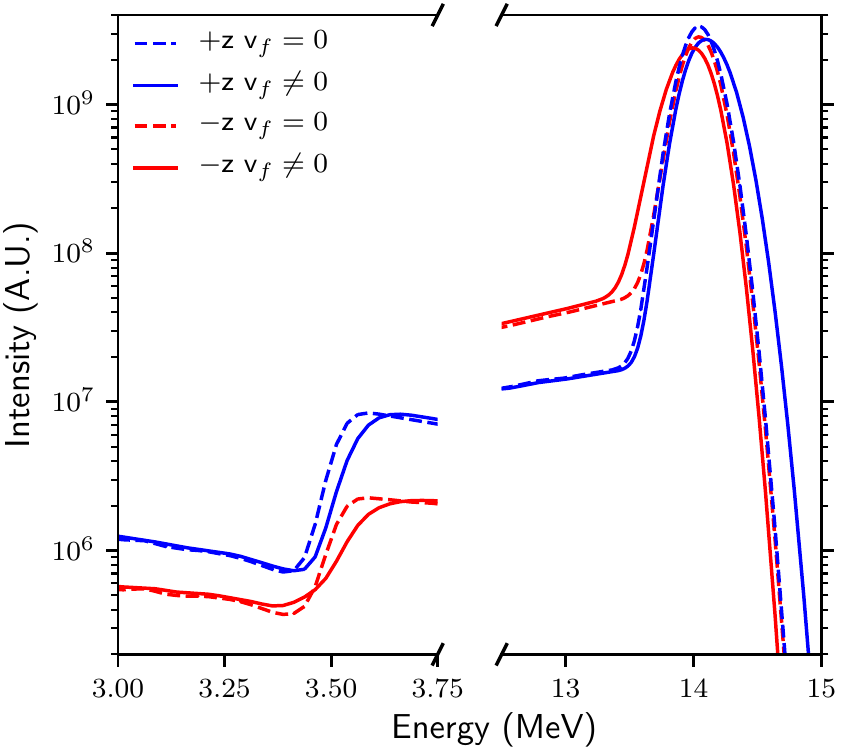}
	\caption{The neutron spectra produced by a P1 perturbed implosion. The +z detector line of sight is anti-parallel to the bulk fluid motion in the hotspot. Spectra along this line of sight are plotted in blue and the anti-podal line of sight in red. The spectra with the effects of fluid motion, $v_f$, neglected are plotted with dashed lines. Shifts and variance changes in the primary spectra are due to hotspot residual kinetic energy. For the backscatter edges, the shifts and variance changes are due to a combination of the hotspot and shell velocities.}
	\label{fig:P1_spec}
	\end{figure}

	The neutron transport was performed with the effects of fluid motion turned on and off to provide a comparison. Sections of the resulting spectra are shown in figure \ref{fig:P1_spec}. The bulk fluid motion has produced the expected shifts of both primary and backscatter edge spectra. The Down Scattered Ratios (DSRs) predict $\rho$R = 0.77 and 0.30 g/cm$^2$ along the +z and -z axes respectively. This areal density difference is mirrored in the intensity of the backscatter edges. An additional areal density measurement can be calculated from the edges \cite{Forrest2012}. \\
		
	\begin{table*}[htbp]
	\begin{tabularx}{\textwidth}{Y  Y  Y  Y  Y  Y }
		\multicolumn{2}{c}{Detector Line of Sight}  & \multicolumn{2}{c}{+z LoS} & \multicolumn{2}{c}{-z LoS} \\
		\hline\hline
		\multicolumn{2}{c}{} & v$_{\mbox{f}} = 0$ & v$_{\mbox{f}} \neq 0$ & v$_{\mbox{f}} = 0$ & v$_{\mbox{f}} \neq 0$ \\ [0.5ex]
		\hline \\ [-1.75ex]
		Primary & Mean (MeV)  & 14.05 & 14.12 & 14.05 & 13.98 \\ [0.5ex]
		
		& Std. Dev. (MeV)  & 0.141 & 0.170 & 0.143 & 0.168 \\[0.5ex]
		
		& $\langle$v$_{\mbox{hotspot}}\rangle$ (km/s)  & - & 130 & - & 130 \\[0.5ex]
		
		& T$_{\mbox{infer}}$ (keV)  & 3.49 & 5.08 & 3.59 & 4.96 \\[0.5ex]
		\hline \\ [-1.75ex]
		nT Edge & Mean (MeV)  & 14.03 & 13.97 & 14.03 & 14.13\\[0.5ex]
		
		Single Fluid Velocity Treatment & Std. Dev. (MeV)  & 0.137 $\pm$ 0.002  & 0.180 $\pm$ 0.001 & 0.135 $\pm$ 0.003 & 0.230 $\pm$ 0.004 \\ [0.5ex]
		
		Model \ref{eqnedge1}& $\langle$v$_{\mbox{shell}}\rangle$ (km/s)  & - & 130 & - & 50 \\[0.5ex]
		
		& T$_{\mbox{infer}}$ (keV)  & 3.30 $\pm$ 0.08 & 5.72 $\pm$ 0.09 & 3.2 $\pm$ 0.2 & 9.3 $\pm$ 0.4 \\[0.5ex]
		\hline \\ [-1.75ex]
		Gaussian Fluid & $\langle$v$_{\mbox{shell}}\rangle$ (km/s)  & - & 130 & - & 50 \\[0.5ex]
		Velocity Treatment \newline Model \ref{eqnedge2} & Std. Dev. v$_{\mbox{shell}}$ / (km/s)  & - & 40 & - & 90 \\[0.5ex]
		
	\end{tabularx}
	\caption{Fitted mean and standard deviations from primary and nT edge spectra. The birth spectrum mean and variance, denoted $a$ and $b^2$, used in the edge fit were initialised using the primary spectra parameters from the antipodal line of sight. An additional linear term was included for the background fit to the edge to account for the nD scattering signal. The ion temperature was inferred using the Brysk variance formula \cite{Brysk1973} and the errors were calculated from the fitting procedure at a fixed shell velocity. For model \ref{eqnedge2}, the fit parameters $a$ and $b$ were held constant, these still take the values obtained from the opposite line of sight primary spectra.}
	\label{table:P1_table}
	\end{table*}

	Fitting to the primary and nT edge spectral shapes was performed and the results are presented in Table \ref{table:P1_table}. The hotspot fluid velocity created a $\sim$ 70 keV mean shift in opposite directions for both the primary spectra. This peak shift corresponds to a neutron-averaged hotspot fluid velocity of $\sim$ 130 km/s. A peak velocity of $\sim$ 500 km/s was found in the hotspot as well as backflow in the +z direction. The magnitude of these velocities are similar to those obtained in a P1 simulation by Spears et al.\cite{Spears2014}. Directly from the Chimera simulation, the burn-weighted line of sight projected velocity was 136 km/s. As the areal density and fluid velocity were increasing through the burn pulse, neutron attenuation means that the spectrally inferred fluid velocity is lowered compared to the burn-weighted velocity. Peak shifts of 113 $\pm$ 16 keV have been observed experimentally on the NIF using polar-direct drive \cite{GatuJohnson2013}, corresponding to bulk fluid velocities of 210 $\pm$ 30 km/s. Similarly in indirect drive experiments\cite{Munro2017}, modest levels of P1 X-ray drive asymmetry create shifts similar to those produced in this simulation. \\
	
	By fitting the edges using a single shell velocity treatment (model \ref{eqnedge1}), neutron-averaged shell velocities of 130 and 50 km/s were found projected along the $-$z and +z axes respectively. Combined with the DSR and hotspot velocity measurements, there is clear diagnosable spectroscopic evidence for a P1 perturbation. The projected fluid velocity seen by neutrons with scattering angles in the range 180$^o\pm$10$^o$ was summed directly within the inverse ray trace. This was used to find a neutron-averaged shell velocity for comparison with that inferred from the backscattered spectra. They were found to be 160 and 78 km/s along the $-$z and +z axes respectively. The higher values are to be expected as high shell velocity regions induce a large positive energy shift on incoming neutrons. These contribute to the signal on the high energy end of the edge fit. Due to the assumption of isotropic centre of mass frame scattering, the fit becomes invalid at energies above the edge where the scattering angles are less than 180$^o$. Including the angular dependence on the differential cross section would allow the fit to be extended and capture the high shell velocity backscattering events. The current fit can be used to set a lower bound on the shell velocities. \\ 
		
	Without fluid velocity effects, the inferred ion temperatures calculated by the edge and the primary spectrum are within $\sim$ 5\% of each other. However the temperatures inferred via the edge are lower. As the intensity of the backscatter edge is proportional to the fuel areal density and the neutron production rate, the time integration introduces an areal density weighting to the nT edge inferred temperature. As the temperature is dropping, the areal density is increasing throughout the burn pulse and thus a lower temperature is expected to be inferred from the backscatter edge. \\
	
	Including fluid velocity effects, variance in the fluid velocity within the hotspot causes broadening in the primary spectra \cite{Murphy2014}. Additional broadening due to the variance in the velocity of the shell is apparent in ion temperatures inferred from the edges. If it is assumed only a single shell velocity is present, a significantly greater apparent T$_{\mbox{i}}$ = 9.30 keV is found from the edge measurement along the $-$z line of sight. This is clearly erroneous when compared to the primary spectral measurement of T$_{\mbox{i}}$ = 5.08 keV, indicting a single shell velocity treatment is invalid. By considering a Gaussian distribution of shell velocities (model \ref{eqnedge2}) and a birth spectrum broadening given by the primary spectrum from the antipodal line of sight (4.96 keV), this larger inferred temperature from the nT edge can be attributed to a 90 km/s variation in the scattering medium velocity. This is analogous to the broadening of the primary spectra due to hotspot velocity variance. Along the $+$z line of sight, lower values of 5.72 keV inferred ion temperature and 40 km/s of fluid velocity variance are found. These shell velocity variances can be explained by considering the differential deceleration of the two sides of the shell. Due to its high inertia, the higher areal density side is decelerating slower hence a lower variation in shell velocity is observed; the opposite is true for the lower areal density side. \\
	
	The nT edge fit is sensitive to the nD single scattering background. With multiple scattering and D(n,2n) included this background smooths out in 1D. As the background includes sources involving multiple interactions, the effect of multi-dimensional perturbations should be lessened. The edge fit will still be applicable in these perturbed scenarios with appropriate background subtraction. Similar background subtraction is performed to extract the DD primary spectrum - in this case a parabolic background fit is used \cite{Hatarik2015}. Thermal motion in the scattering medium was not considered here but will add additional broadening to the edge in experimental results. 
		
	\subsection{Areal Density Asymmetry Effects on DSR Region}

\begin{figure}[htp]
		\centering
		\includegraphics*[width=0.495\textwidth]{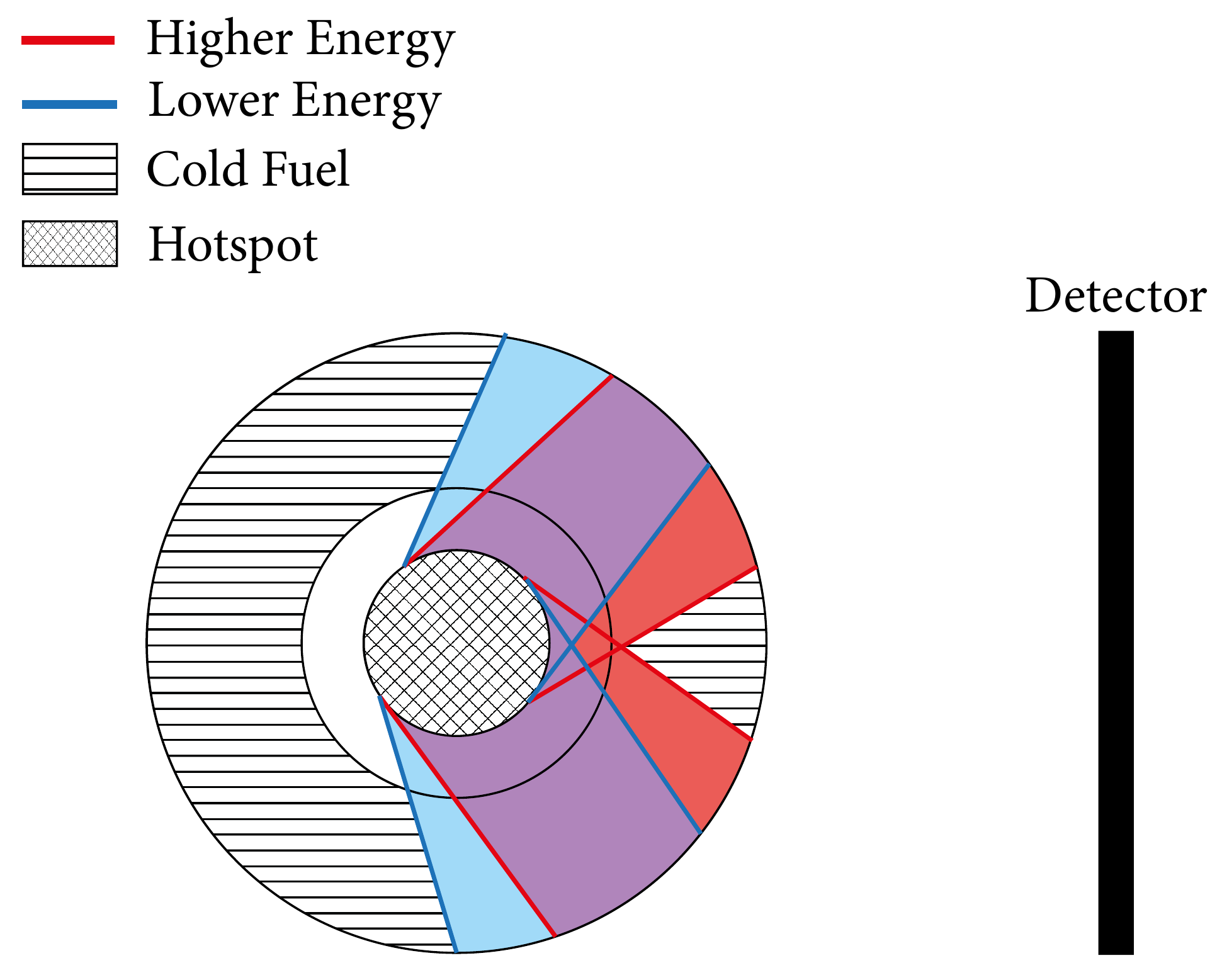}
		\caption{Diagram showing two scattering cones for two different energy gates. Scale of hotspot and cold fuel has been altered to emphasise the scattering cone geometry. Due to an extended source there is overlap between the available scattering regions. Higher energy scattered neutrons will sample regions at a lower angle to the detector line of sight. The reverse is true for lower energy scattered neutrons. For a 14 MeV neutron the scattering angles from D (T) down to 10 and 12 MeV are 47.5$^o$ (59.5$^o$) and 31.9$^o$ (39.5$^o$) respectively.}
		\label{fig:coneangle}
	\end{figure}

	The degree of down scatter of DT fusion neutrons is often used as a measure of areal density. This is done through a DSR measurement which is converted to an areal density through a fit to 1D neutron spectra \cite{Johnson2012,Frenje2013}. When considering 3D effects, the energy-angle relation given in equation \ref{elastic} is a vital reference. The areal density traversed by the downscattering neutrons is at an angle to the line of sight \cite{Johnson2012}. For a given energy range, this defines a cone of scattering over which a neutron-averaged areal density is measured, see figure \ref{fig:coneangle}. The line of sight areal density is only experienced by the un-scattered or low scattering angle neutrons. This draws a distinction between areal density measurements from primary neutrons, such as the FNADs, and scattered neutrons, such as the nTOFs. Energy gating and spectra can use this geometric feature of elastic scattering to measure different spatial regions of the implosion.\\ 
	
	Variations in plasma conditions may influence the slope of the scattered spectrum. The shape of down scatter spectrum is set by the differential cross section, the energy distribution of the scattering neutrons and the areal density of the scattering medium. Concentrating on the last of these, if the areal density is greater at higher scattering angles then the flux of lower energy scattered neutrons will be increased. As the spectrum down to 10 MeV is dominated by elastic scattering from DT, changes of spectral shape in this region could indicate fuel areal density asymmetry. \\
	
\begin{figure}[htp]
		\centering
		\includegraphics*[width=0.40\textwidth]{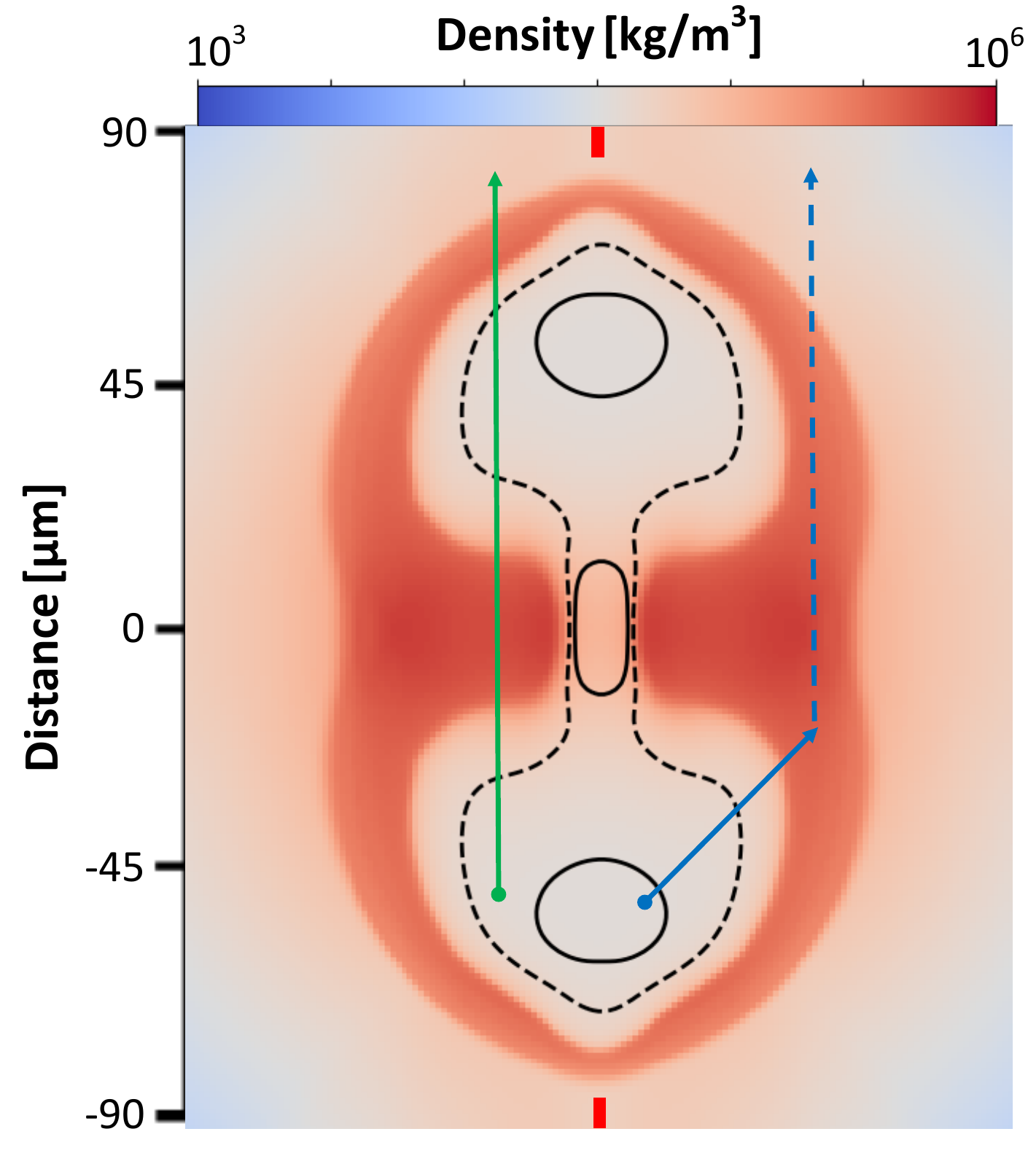}
		\caption{The density in the x-z plane at peak neutron production. The black contours show the single (solid line) and double e-folding (dashed line) of the unattenuated primary neutron fluence. The green line shows the path of an un-scattered neutron to a polar detector. The path of a singly-scattered neutron is shown by the blue lines. The probability of scattering is proportional to the areal density along the solid blue line. The axis of symmetry is on the line connecting the vertical red ticks.}
		\label{fig:P2_dens_RR}
	\end{figure}

	To illustrate this effect, the neutron spectrum for an implosion with a P2 drive asymmetry will be considered, see figure \ref{fig:P2_dens_RR}. The magnitude of the drive asymmetry, 3\% P2/P0, used in the Chimera simulation was such that the neutron yield was close to experimental values (3.19 $\times$ 10$^{15}$ without alpha-heating). The positive P2 asymmetry created a larger $\rho R$ along the waist compared to the poles. However, due to the hotspot being greatly distorted, the areal density experienced on average by a neutron leaving the capsule differs significantly from the $\rho R$ taken from the simulation centre point.\\ 
	
\begin{figure}[htp]
		\centering
		\includegraphics*[width=0.495\textwidth]{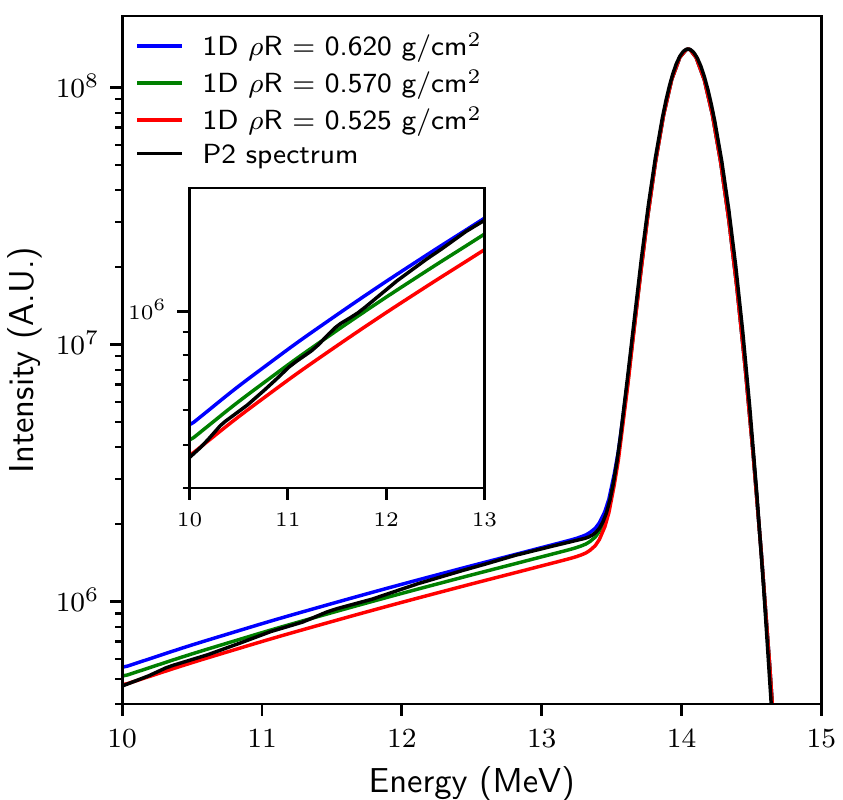}
		\caption{The neutron spectrum produced along the axis of symmetry by a capsule implosion with an imposed P2 perturbation, black line. The blue, green and red lines show the single downscatter spectrum for 3 different areal densities. The areal densities used in descending order are the line of sight neutron-averaged areal density, the inferred areal density from the DSR of the P2 simulation and an areal density chosen such that 1D spectrum matches that P2 spectrum at 10 MeV. Inset is a zoomed in plot over the 10-13 MeV region.}
		\label{fig:DSR_grad}
	\end{figure}

	In order to infer the change in areal density with angle from neutron spectra, it was assumed that the primary spectrum measured and the spectrum of neutrons to be downscattered were the same. Thus by matching the measured primary spectrum, a set of calculated 1D neutron spectra would match the downscattered spectrum from the perturbed implosion when the areal densities were equal. Various 1D isobaric hotspot simulations were performed where the temperature and velocities were tailored to fit the Brysk cumulants of the primary spectrum from the P2 simulation. Due to the low neutron-averaged areal densities, the single scatter approximation used in the neutron transport is valid. For a detector viewing along the axis of symmetry, the large areal density surplus at the waist is avoided by neutrons born in the polar jets which scatter into the detector line of sight, c.f. solid blue line in figure \ref{fig:P2_dens_RR}. From the simulation data, it is seen that, as the scattering angle is increased from 30$^o$ to 60$^o$, the neutron-averaged areal density along scattering paths decreases. Intuitively this will increase the slope in the 10 - 12 MeV range of the spectrum. Figure \ref{fig:DSR_grad} shows the resultant neutron spectrum from the P2 simulation with a polar detector. \\
	
	The P2 spectrum DSR infers a $\rho R = 0.57$ g/cm$^2$; this areal density represents the weighted sum over all scattering paths exemplified by the solid blue line in figure \ref{fig:P2_dens_RR}. The $\langle \rho R\rangle$ along the detector direction was found to be 0.62 g/cm$^2$. The 1D spectra with these areal densities intersect the P2 spectrum at different energies, see figure \ref{fig:DSR_grad}. A 1D simulation with $\rho R = 0.525$ g/cm$^2$ intersects close to a neutron energy of 10 MeV. This corresponds to a 15\% change in $\rho R$ between the line of sight and the angular range near 10 MeV. Minotaur was used to construct single-scattering $\rho R$-DSR relationships for the 10--11, 11--12 and 12--13 MeV regions. Applying these to the P2 neutron spectrum, $\rho R_{\mbox{10-11}} =$ 0.56 g/cm$^2$, $\rho R_{\mbox{11-12}} =$ 0.58 g/cm$^2$ and $\rho R_{\mbox{12-13}} =$ 0.60 g/cm$^2$ were calculated. The average angles in these ranges are $49^{\circ}$, $41^{\circ}$ and $31^{\circ}$ respectively. These were weighted by the differential cross section with the assumption of a 50:50 mixture of DT. In this way spectra can be utilised to calculate the magnitude of neutron-averaged areal density asymmetry. As each energy range includes integration over the hotspot volume and scattering cone, the areal density from simulation centre is not accessible. At bang time the centre point areal densities are 0.4 and 2.2 g/cm$^2$ along the pole and across the waist respectively. This asymmetry is significantly larger than that inferred from the neutron spectrum.\\
		
	\section{Neutron Imaging}
	
	Hotspot and cold fuel shell shape capture the spatial deviations from symmetry seeded by various perturbation sources. The form of the areal density asymmetries can identify the failure mechanism. Within this section, the degree to which these asymmetries affect the measured neutron images will be investigated using synthetic images.
	
	\subsection{Primary Images}
	
\begin{figure}[htp]
	\centering
	\includegraphics*[width=0.495\textwidth]{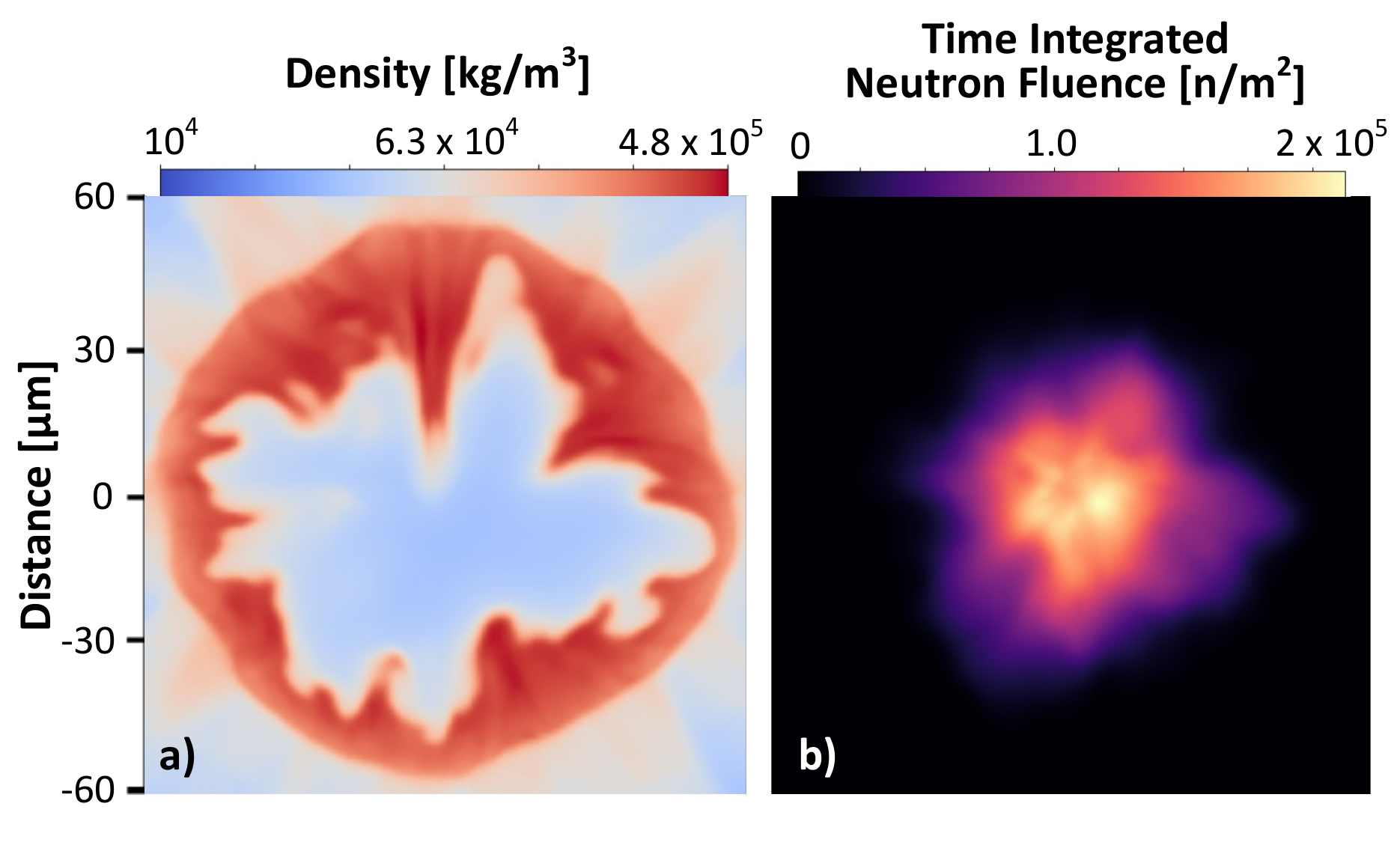}
	\caption{\textbf{a)} 2D y-z density slice at neutron bang time through a 3D Chimera simulation with randomly seeded Rayleigh-Taylor velocity perturbations applied \cite{Walsh2017}. The magnitude of the perturbations were set such that the neutron yield was close to experimental values (4.80 $\times$ 10$^{15}$ without alpha-heating). \textbf{b)} A time integrated primary neutron image down the x-axis, through the density slice given.}
	\label{fig:multimode}
	\end{figure}

	Primary neutron images map out a set of line integrals of D(T,n) reaction rate with an attenuation factor due to scattering and other interactions. These line integrals can be easily solved via the inverse ray trace method outlined in section II. Primary neutron images can be used to outline the shape of the hotspot, as seen in figure \ref{fig:multimode}. The low density limbs seen in the density slice are not present in the primary neutron image due to the increased conductive losses in these regions. \\

\begin{figure}[htp]
	\centering
	\includegraphics*[width=0.495\textwidth]{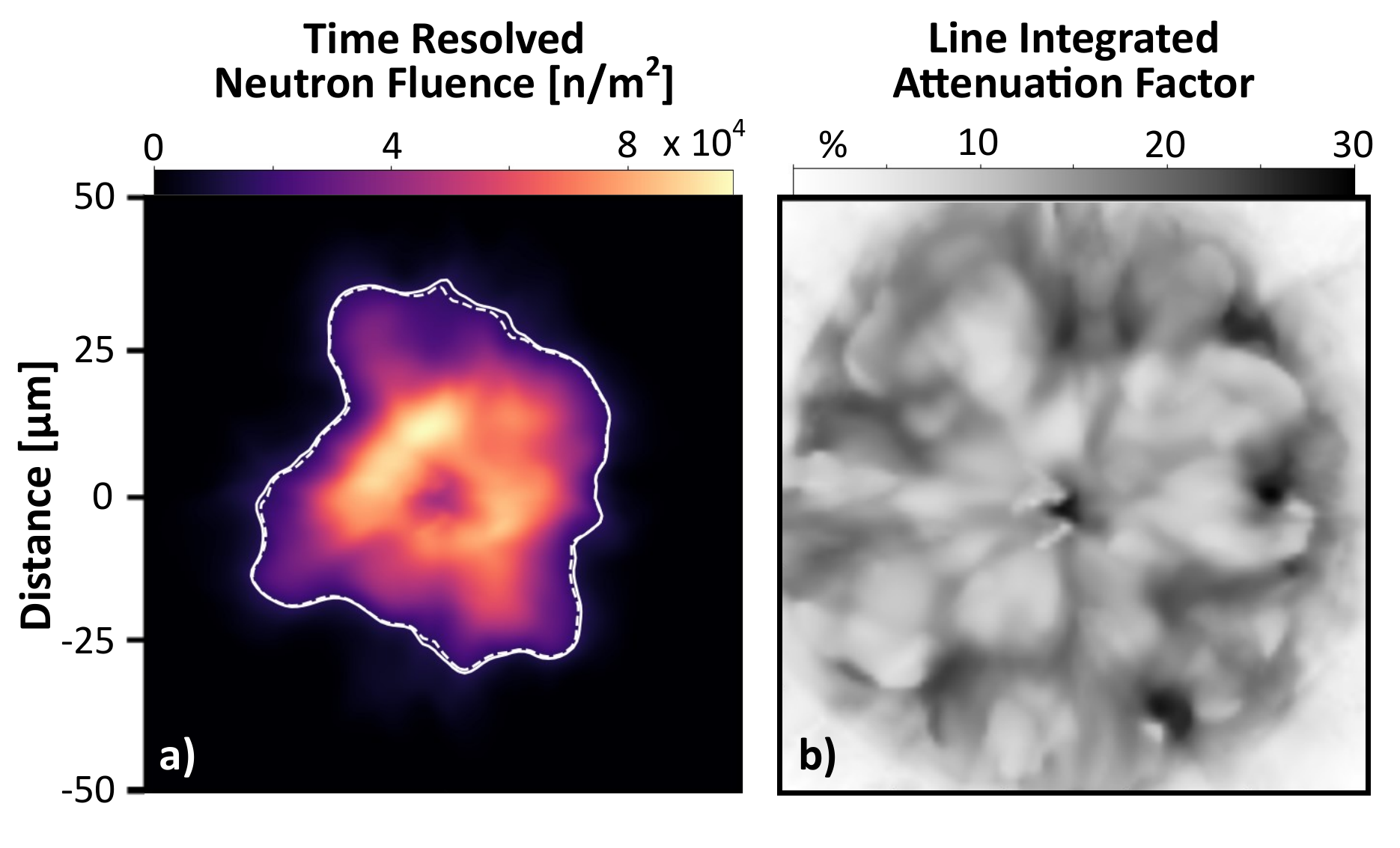}
	\caption{Large areal density differences can create variation in the level of attenuation for primary neutrons. Using the same simulation as used in figure \ref{fig:multimode}, \textbf{a)} shows the attenuated primary neutron image down the z-axis at bang time with 1$\mu$m resolution. The white lines map out the 17\% contour which is used to measure hotspot shape \cite{Grim2013}. The solid line is the contour when attenuation is neglected while for the dashed line attenuation is included. \textbf{b)} shows the spatial variation in the level of attenuation. This is defined through a line integrated attenuation factor given by one minus the ratio of the attenuated to the unattenuated flux at the detector.}
	\label{fig:attenuation}
	\end{figure}

	Through multiple lines of sight, tomographic techniques can be used to form a 3D reconstruction of the neutron reaction rate \cite{Volegov2015,Volegov2017}. Attenuation from increased areal density along a given line of sight leads to reduction in measured signal. This can cause misidentification of darker regions as regions of low production if attenuation effects are ignored. For more symmetric implosions the attenuation would lead to a limb darkening effect. This will reduce the inferred size of the hotspot if intensity contours are used. However, this effect is likely to be marginal, as the multimode simulation shows in figure \ref{fig:attenuation}. When imaged at 1$\mu$m resolution, the level of attenuation is observed to vary by at most 20\% within the hotspot. The shape inferred via the 17\% contour with attenuation effects on and off is within 2.5$\mu$m at all points. At 5$\mu$m resolution, the variation in attenuation is reduced to below 13\%. Thus the error induced by neglecting attenuation in analysis of primary images is tolerable for high mode and small amplitude low mode perturbations. \\
	
	\subsection{Scattering Medium Density Analysis}
	
	Scattered neutron images can be used to identify areas of increased density as this will lead to increased scattering. However, these images contain information about both the density and the neutron flux at the scattering site. Therefore additional analysis is required to infer the scattering ion density. \\
	
	Fluence compensation is a technique which can be used to approximately decouple the primary flux from scattered neutron images and therefore directly image the scatterer density \cite{Casey2016}. The technique involves using the primary image to infer the primary fluence at the scattering locations. To do this an average scattering angle is used and the emission rate is assumed to be constant along the line of sight. Using tomographic techniques the constant emission rate assumption can be discarded; however, this will not be explored here. The averaged angle is calculated using the neutron spectrum found along the same line of sight. This is done by calculating the average angle of scattering for a 14 MeV neutron with a given outgoing neutron energy using the D and T differential scattering cross sections. This angle is then averaged using the neutron flux from the spectrum at the outgoing energy. Dividing the scattered neutron image by the approximated primary fluence retrieves the product of the areal density seen by these scattered neutrons and the differential cross section. Attenuation effects on the primary fluence calculation are neglected, see section IV A for discussion. \\
	
	\begin{figure}[htp]
	\centering
	\includegraphics*[width=0.495\textwidth]{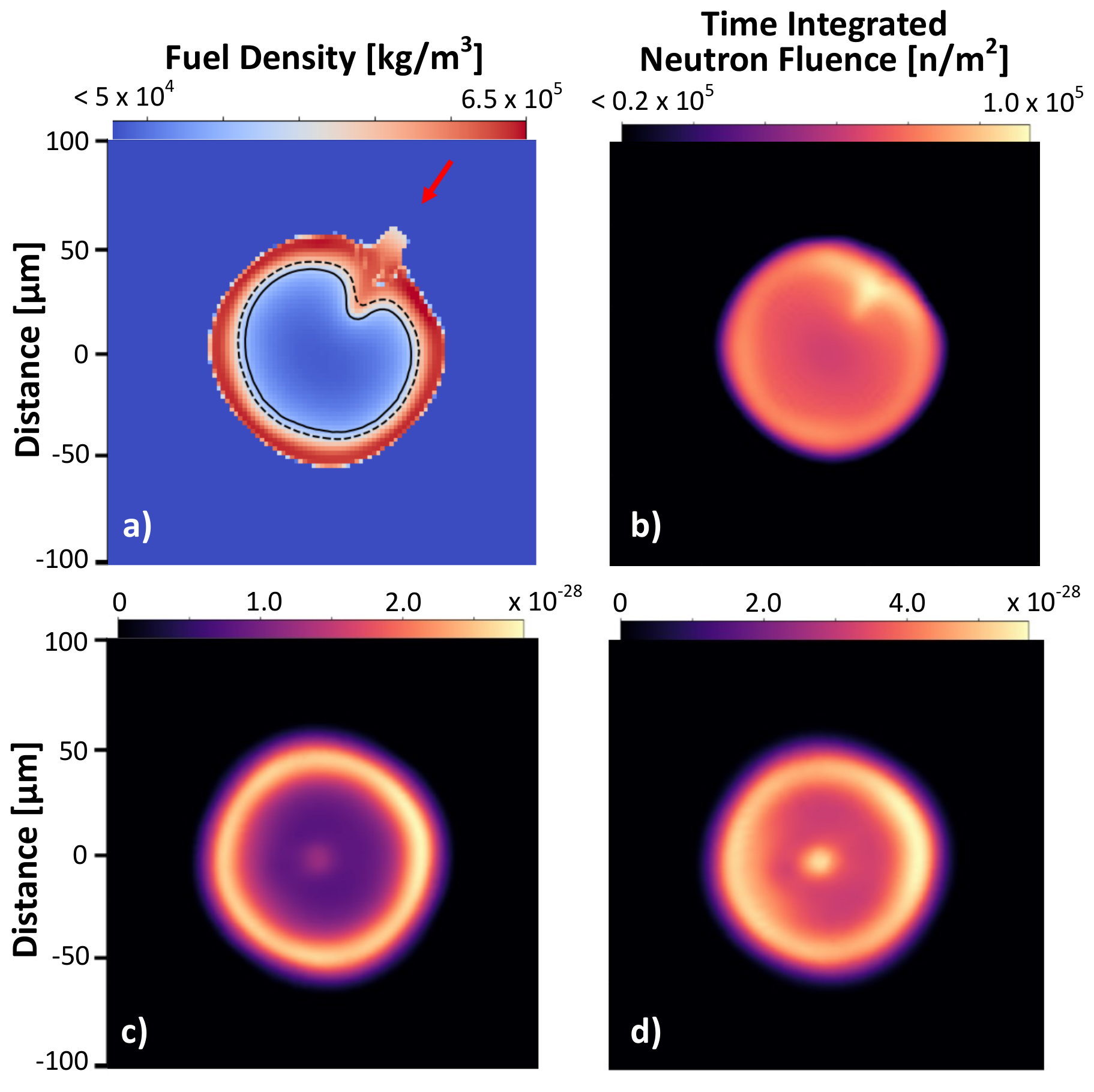}
	\caption{\textbf{a)} Fuel density slice in the x-y plane at bang time and the fusion reaction rate contours overlaid. The solid and dashed lines represent a single and double e-folding of the reaction rate respectively. \textbf{b)} A scattered neutron image taken down the $z$-axis with a 10-12 MeV gate. \textbf{c)} and \textbf{d)} Fluence compensated images showing the single spike perturbation along the detector line of sight (90$^o$,45$^o$), shown by the red arrow in \textbf{a)}. The two images correspond to two energy gates, 10.0-11.5 MeV and 11.5-13 MeV respectively. The average scattering angles inferred from the neutron spectrum are $\sim$ 45$^o$ and 30$^o$ respectively.}
	\label{fig:fluencecomp}
	\end{figure}
	
	Different energy gates can be used to sample various regions of the dense shell, see figure \ref{fig:coneangle}. To illustrate this technique we will consider a calculation where a single spike Rayleigh-Taylor\cite{Taylor2013b} velocity perturbation has been applied to an otherwise symmetric implosion at peak velocity. At initialisation, the maximal radial velocity of this spike was a factor of $\sim$ 2 times greater than in the rest of the shell. Using two energy gates above 10 MeV allowed separate imaging of the spike density and the shell density at higher angles, see figure \ref{fig:fluencecomp}. The time dispersion of the neutrons would allow this technique to be applied along a single line of sight. By viewing the spike along its axis, blurring due to radial motion is avoided in the time integrated image. A time integrated scattered neutron image perpendicular to the spike axis reproduces the shape of the perturbation.\\

	The 11.5-13 MeV image can be used to estimate the areal density of the cold fuel spike and the preceding shell. Dividing out by the differential cross section evaluated at the average angle of 30$^o$, an areal density of $\sim 2.0$ g/cm$^2$ was obtained. An areal density of 1.8 g/cm$^2$ was found at bang time directly from the Chimera simulation.
	
	\subsection{Flange-Mounted Neutron Activation Diagnostic System (FNADs)}

	The FNADs measure the incident neutron fluence modulated by the activation cross section which possesses an energy threshold \cite{Bleuel2012,Yeamans2012,Yeamans2017}. This system is distributed over $4\pi$ of solid angle and aims to capture the 3D nature of the areal density. Since the neutron fluence is attenuated by areal density, variation in FNAD signal is correlated with capsule areal density variations. The hotspot is generally extended and therefore it is expected that the areal density variation inferred in this way is correlated with the neutron-averaged areal density. Due the energy dependence of the activation cross section, the Doppler shifts induced by fluid velocity in the hotspot will also affect the FNADs signal \cite{Bleuel2012,Rinderknecht2018}. Flows towards an activation sample will increase the signal measured.\\
		
\begin{figure}[htb]
	\centering
	\includegraphics[width=\columnwidth]{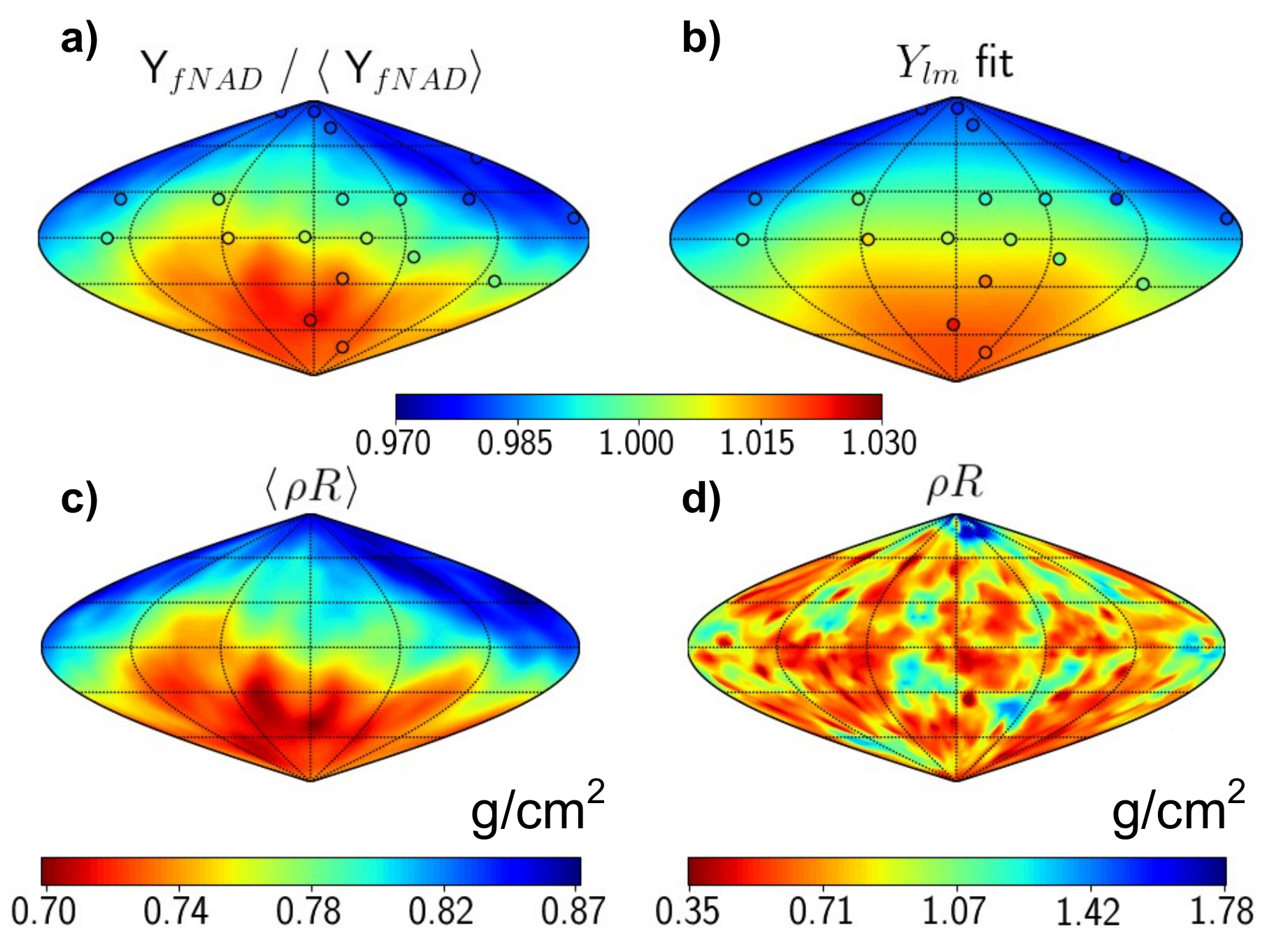}
	\caption{Sky maps of \textbf{a)} the synthetic FNAD signal for an implosion with multimode perturbations applied; \textbf{b)} a $l, m \leq 2$ spherical harmonic fit performed on the synthetic FNAD data sampled at the experimental detector positions; \textbf{c)} the neutron-averaged areal density and \textbf{d)} the line integrated density from the simulation centre point weighted by the burn history.}
	\label{fig:FNAD}
	\end{figure}

	To form the synthetic Zr$^{90}$ FNADs a forward ray trace was performed from all the emitting cells along a detector direction vector. A birth spectrum was created for each emitting cell based on the temperature and fluid velocity. The areal density along the neutron path was used to attenuate the number of neutrons measured. Using the 3D multimode simulation shown in figure \ref{fig:multimode}, 5000 synthetic FNADs were distributed at uniform angular intervals over 4$\pi$. The resultant measurement is given in figure \ref{fig:FNAD} a). A low mode ($l, m \leq 2$) spherical harmonic fit was performed on the 19 experimental FNAD locations \cite{Yeamans2012,Yeamans2017}, shown as coloured circles, showing asymmetry in both the polar, $l$, modes and azimuthal, $m$, modes, see figure \ref{fig:FNAD} b). \\
	
	The degree of smoothing of high mode perturbations can be seen by comparing the neutron-averaged areal density (figure 14c) and the line integrated density from the simulation centre point weighted by the burn history (figure 14d). As expected, the neutron-averaged areal density, as defined in section II, shows the same features as the FNAD map. However, the map of centre point $\rho$R weighted by the burn history is notably different. In the $\langle\rho$R$\rangle$ map, the high mode detail has been greatly smoothed and the magnitude of the resulting low mode alias is lower. The extended nature of the hotspot means that the bubbles and spikes seen clearly in the $\rho$R map are sampled from multiple starting points. Spikes surrounded by emitting material thus appear spatially larger. Also, when viewed down the spike axis, the attenuation effect appears reduced due to emitting plasma surrounding the spike. Hence high mode perturbations are not resolvable, no matter the number of activation detectors, and the magnitude of the low mode sample of these perturbations is lowered. Experimental FNADs sky-maps often exhibit modes $l,m \leq 2$ of similar or larger magnitude as those found in the multimode synthetic data \cite{Bleuel2012,Hurricane2016}. In this multimode simulation a large polar spike produces a P1 signal in the FNADs. Possible high mode detail missed by the FNADs would be visible in well-resolved primary and scattered neutron images, e.g. figures \ref{fig:multimode} and \ref{fig:attenuation}. The neutron-averaged fluid velocity inferred from the primary spectra (29 km/s) accounts for $\approx$ 25$\%$ of the FNAD variation. This L=1 effect exacerbates the aliasing of the high mode areal density variation, although correction of the fluid velocity effect is possible with a spectroscopic hotspot velocity measurement\cite{Rinderknecht2018}. Performing downscattered neutron spectral measurements along the poles, the $\rho R_{\mbox{DSR}}$ were found to be 0.71 and 0.81 g/cm$^2$ respectively. These are consistent with the P1 measured in the FNADs. The DSR measurement samples between $\sim 30-60$ degrees away from the detector axis, on figure \ref{fig:FNAD} c this corresponds to the area between the two lines of latitude nearest the detector location.\\
	
	\section{$\gamma$-Ray Diagnostics}
	
	Time histories of the 16.7 MeV $\gamma$-rays produced by a small fraction, $\sim$ 4$\times$10$^{-5}$, of DT fusion reactions can be used to measure the bang time and burn width \cite{Mack2006,Kim2012}. Nuclear interactions between fusion neutrons and nuclei can also produce $\gamma$-rays. Notably, neutrons inelastically scattering from carbon in the ablator produce a 4.4 MeV $\gamma$-ray \cite{ENDF}. This signal provides a functional diagnostic of ablator areal density \cite{Cerjan2015}. Spatial and temporal $\gamma$-ray diagnostics could therefore be used to measure ablator areal density asymmetry and evolution. It should be noted that imaging of 4.4 MeV $\gamma$-rays has yet to be achieved at the NIF.
	
	\subsection{Imaging the Carbon Ablator via Inelastic Neutron Scattering}
	
	Cold fuel spikes entering the hotspot cause increased conductive losses; these spikes can be seeded by perturbations within the ablator. It is therefore critical to image the shape of the ablator near peak neutron production. By imaging the C$(n,n_1 \gamma)$ $\gamma$-rays, the position of ablator around bang time can be seen. Since the C$(n,n_1 \gamma)$ cross section is a relatively weak function of energy in the 6-20 MeV range, only scattering of the primary DT neutrons was considered as this will provide the dominant signal. On the path out of the hotspot and through the cold fuel layer, this primary neutron flux will be diluted by $r^2$ and a small fraction scattered away. Therefore spikes of ablator close to the hotspot will be highlighted due to the larger neutron fluence.  \\
	
\begin{figure}[ht]
	\centering
	\includegraphics*[width=0.495\textwidth]{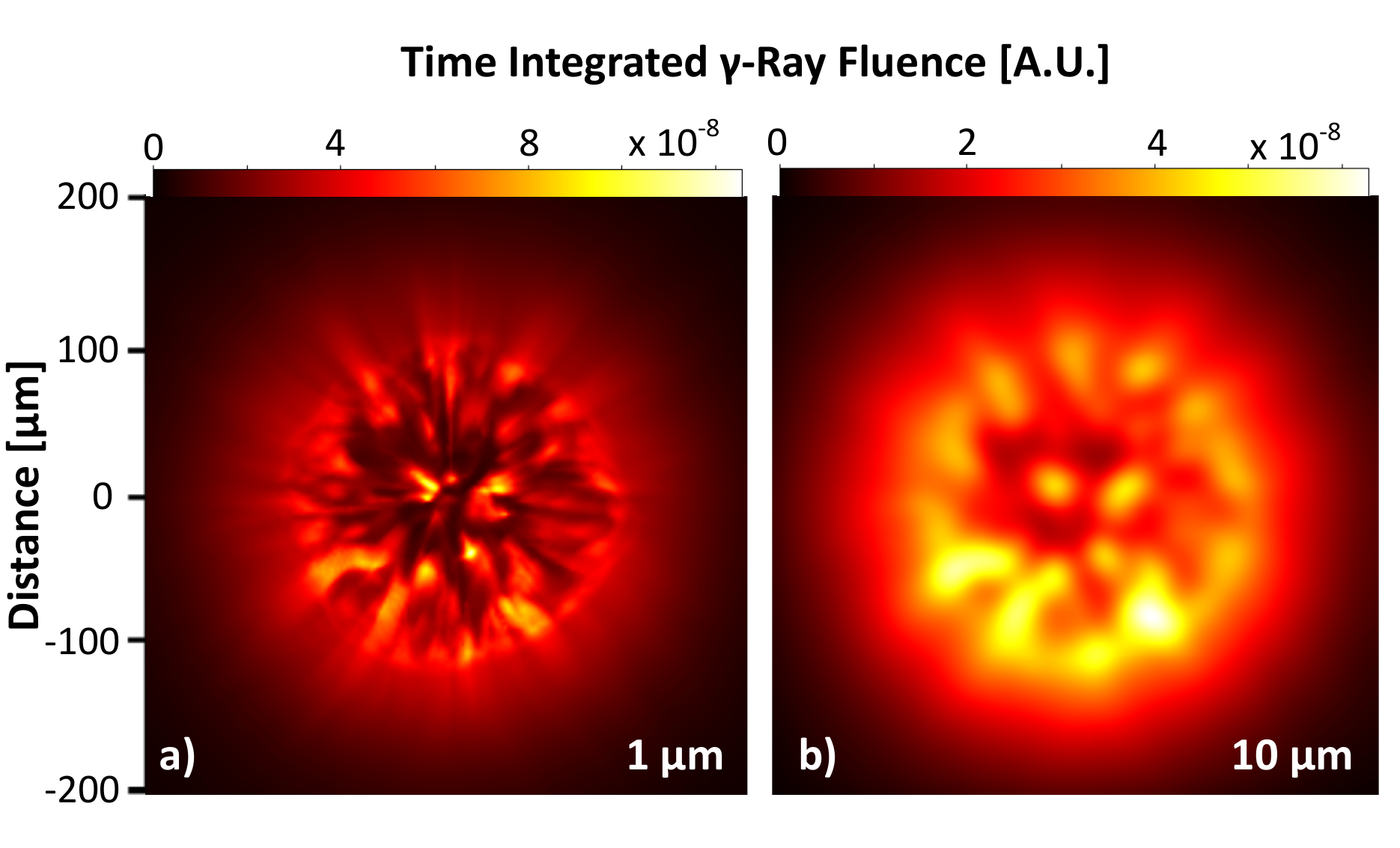}
	\caption{\textbf{a)} Time integrated carbon $\gamma$-ray image at the resolution of the simulation (1 $\mu$m) for multimode simulation. \textbf{b)} A 10 $\mu$m Gaussian filter applied to image \textbf{a)}. Carbon spikes from all 4$\pi$ of solid angle and the blow-off region are seen.}
	\label{fig:cgamma}
	\end{figure}

	Time integrated $\gamma$-ray images were created using the multimode simulation also used in section IV.  The spikes created by the perturbations and the blow-off region are clearly visible in figure \ref{fig:cgamma}. The presence of spikes still persists after the addition of a 10$\mu$m Gaussian filter. Due to the low signal, previous Monte Carlo $\gamma$-ray images have exhibited poor statistics \cite{Bradley2003}. The inverse ray trace method has led to a clearer image of the ablator material.
	
	\subsection{Fusion and Carbon $\gamma$-ray Histories}

	DT fusion $\gamma$-rays can be used to track the progress of burn in time without complications from additional interactions in flight to the detector. Without significant alpha heating, the areal density is still increasing throughout the fusing period and peak compression is reached after bang time. Thus it is expected that the peak carbon $\gamma$ signal will be reached after bang time for a symmetric case. Areal density asymmetries which disrupt the hotspot will alter both the fusion and carbon $\gamma$-ray histories. As the time between peak neutron production and peak areal density increases, an increasing delay between fusion and carbon $\gamma$-ray peaks will be observed.

	\begin{figure}[ht]
	\centering
	\includegraphics*[width=0.495\textwidth]{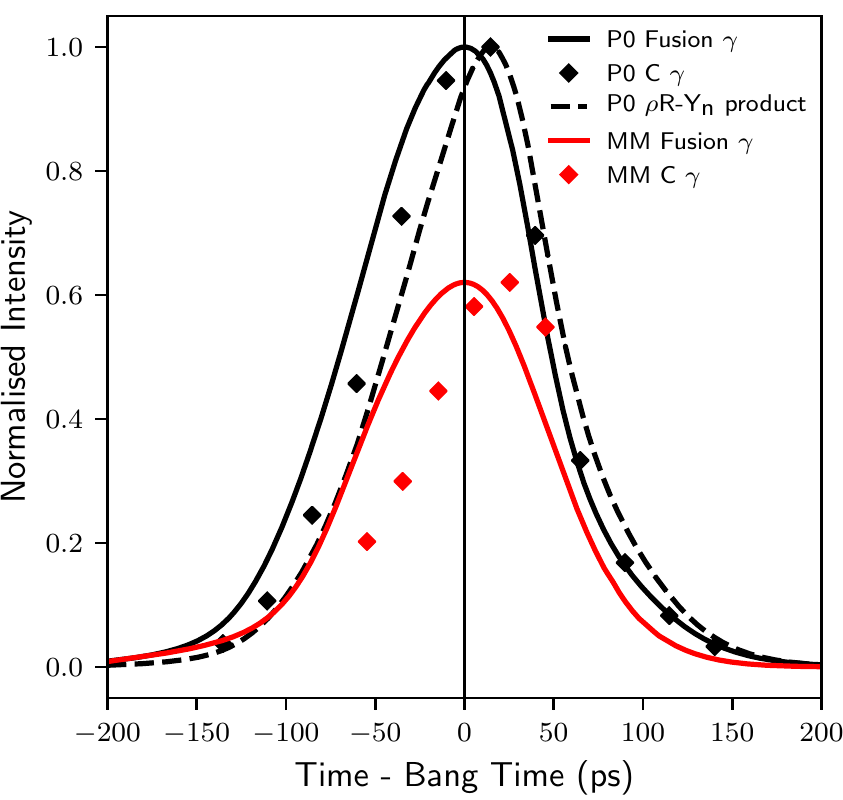}
	\caption{Time histories of D$(\mbox{T},\gamma)$ and C$(n,n_1 \gamma)$ yields for a symmetric implosion (P0) and the multimode simulation (MM). It is observed that in both cases the peak carbon $\gamma$ signal was reached after bang time. For the multimode simulation, a more significant time shift in carbon $\gamma$ signal is observed. The fusion $\gamma$ signals are normalised to the P0 peak value, the carbon $\gamma$ signals peaks match their respective fusion $\gamma$ signals. The dashed line shows the evolution of the product of ablator areal density and fusion reaction rate in the symmetric simulation.}
	\label{fig:cgammatime}
	\end{figure}

	Figure \ref{fig:cgammatime} presents the $\gamma$-ray histories from the multimode case and the symmetric simulation from which the multimode simulation was initialised. For the symmetric case, peak carbon $\gamma$ signal occurred $\sim$ 15 ps after bang time. \\
	
	The time difference between peak fusion and carbon $\gamma$ production was 30 ps for the multimode case. This was due to the spikes of dense DT shell entering the hotspot and cooling it. The increased conductive losses caused burn truncation and a drop in hotspot pressure, leading to an early bang time and a delayed peak compression. The source of carbon $\gamma$-rays given by the ablator entrained behind the spikes and remaining unperturbed ablator continued to converge throughout neutron production. \\
	
	In the symmetric case, the time evolution of the carbon-$\gamma$ signal should correlate with the product of the ablator areal density and the fusion reaction rate, up to extended hotspot and attenuation effects. Peak ablator areal density occurs around 100 ps after bang time. Hence attenuation by the cold shell will increase throughout the burn pulse. Therefore the simulated carbon $\gamma$ signal is expected to be increasingly reduced with increasing time compared to the product of ablator areal density and burn rate. This reduction in carbon $\gamma$ signal is compounded by the hotspot radius decreasing with time causing the mean neutron path through the ablator to be reduced. In reality a fraction of the neutrons scattered prior to exiting the fuel will inelastically scatter off the ablator producing $\gamma$-rays which the simulation does not account for. Hence the real signal is expected to lie between the extremes of no attenuation with a point source, given by the areal density burn rate product, and full attenuation with an extended source, given by the inverse ray trace model.
	
	\section{Conclusions}
	Synthetic nuclear diagnostics were used to explore novel approaches to unfolding hotspot, dense DT shell and ablator properties. These techniques could extend the capabilities of existing experimental diagnostics and can be used to determine requirements for future diagnostic development. \\
	
	Novel functional fits based on a simplified backscatter model were presented and used to infer scattering medium properties. With the use of anti-podal primary and nT edge spectral measurements, this model inferred birth spectral moments and shell velocity means and variances. Accurate measurement of the backscatter edge requires fine energy resolution, $\lesssim$ 10 keV, around the edge due to its sharp gradient. This need has recently been answered by the development of Cherenkov detectors at the NIF \cite{Schlossberg2017}. These will increase the precision of neutron spectral fits performed on future experimental data.\\
	
	The DSR range in the spectrum can exhibit significant slope changes in the presence of areal density asymmetries. By comparison to 1D spectra, areal densities within different angular ranges were calculated for a P2 simulation. Experimental spectral data measuring down to 10 MeV exists\cite{Hatarik2015} and a comparison of inferred areal densities, FNADs and neutron images would test the viability of this analysis. Measuring the primary spectrum, backscatter edges and DSR range, and hence the inferred hotspot and shell velocities and areal density asymmetries, would allow for a more complete description of low mode perturbations.  \\
	
	Multiple energy-gated neutron images and the fluence compensation technique allow a detailed description of shell conditions. Multiple energy gates probe different angular ranges and hence localised features such as high mode spikes and the tent scar can be isolated in this scheme. For a simulation with a single spike perturbation, fluence compensated images faithfully recovered the areal density conditions near bang time. Performing the gating on a single line of sight would allow such features to be imaged with the same projection characteristics. The recent increase in neutron yields \cite{Meezan2015,Edwards2016} would allow for sufficient signal-to-noise ratio in narrower energy gates.\\
	
	The synthetic FNAD signal showed strong correlation with the neutron-averaged areal density. Therefore both the hotspot shape and shell areal density asymmetries contributed to the measured signal. For a multimode simulation, the high mode perturbations were not resolvable by the FNADs due to the extended nature of the hotspot. Comparison with additional imaging techniques is essential in drawing the distinction between true and aliased low mode asymmetries inferred from the FNADs. Hotspot velocity appears as an additional low mode signal, further reducing the ability to measure high mode areal density variation. \\
	
	Carbon-$\gamma$ imaging reveals spatial variation of the ablator areal density. For a multimode simulation, a spatial resolution of 10 $\mu$m on a time integrated image was sufficient to observe the perturbed structure. For the same simulation, the peak carbon $\gamma$ signal was found to be delayed compared to a symmetric implosion. Observation of a delayed signal compared to the fusion $\gamma$ peak is indicative of burn truncation. In this case, the truncation was caused by high mode perturbations causing increased conductive losses in the hotspot plasma around bang time. The analysis in section V B suggests a time resolution of $\lesssim$ 15 ps would be needed to resolve this peak shift difference.\\
	
	Development of synthetic X-ray diagnostics alongside the nuclear diagnostics developed here is key to investigating experimental data trends with hydrodynamic simulations. Fitting to the high energy X-ray spectra has been used to infer electron temperatures \cite{Jarrott2016}; these temperatures are not sensitive to bulk fluid velocities. Hence comparison to inferred ion temperatures from neutron spectra could be used to infer ion-electron temperature differences and bulk fluid velocity variance. Mix of ablator material into the hotspot is attributed to distinctive features in X-ray images and spectra \cite{Ma2017,Regan2012}. Correlation with neutron and carbon $\gamma$ images can be used to find the effect of this mix on hotspot shape. The inverse ray trace method is being adapted to perform radiation transfer of X-rays to create images and detailed spectra. 
	
	\section*{Acknowledgements}
	The authors would like to thank Dr. Brian Spears, Dr. Dan Casey, the nToF team, Dr. David Fittinghoff, Dr. Gary Grim and Dr. Prav Patel of the Lawrence Livermore National Laboratory for many invaluable discussions. The authors would also like to thank the referees for their detailed comments and constructive feedback. The results reported in this paper were obtained using the UK National Supercomputing Service ARCHER and the Imperial College High Performance Computer Cx1. This work was supported by the Lawrence Livermore National Laboratory through the Academic Partnership Program, the Engineering and Physical Sciences Research Council through Grant Nos. EP/P010288/1 and EP/M011534/1, and by AWE Aldermaston.
	
	\appendix
	\section{Minotaur - 1D Discrete Ordinates Code}

	This appendix contains description of the numerical schemes employed with Minotaur and presents the effects of multiple scattering and the error induced in the single scatter approximation.\\

	The neutron transport equation can be expressed in the form of the Boltzmann equation as the nuclear interaction length is short:
	\begin{align}\label{ntrans}
	&\left[\frac{1}{v}\frac{\partial}{\partial t}+\hat{\Omega}\cdot\vec{\nabla}+n(\vec{r},t)\sigma(E)\right]
	\psi(\vec{r},\hat{\Omega},E,t) = \\ &S_{ex}(\vec{r},\hat{\Omega},E,t)\ \nonumber +\\ &\int_0^{\infty} dE'\int d\hat{\Omega}' n(\vec{r},t)\sigma_s(\hat{\Omega}'\cdot\hat{\Omega},E' \rightarrow E)\psi(\vec{r},\hat{\Omega}',E',t) \nonumber
	\end{align}
	Where $\psi$ is the angular neutron flux and $v$ and $\hat{\Omega}$ are the speed and direction of motion of the neutrons. The number density of the interacting species, $n$, has total and double differential cross sections denoted by $\sigma$ and $\sigma_s$. Sums over different species are implicit. The time independent case will be considered as discussed in section II. \\
	
	The effect of neutron-ion interactions on the neutron distribution function are described through double differential cross sections, shown in the integral source term on the RHS of equation \ref{ntrans}. The summation of cross sections for these interactions, as well as absorption, retrieves the total cross section which appears in the streaming operator on the LHS. This operator describes free-streaming of the neutrons with attenuation from the total interaction cross section.\\
	
	Neutron transport including all relevant nuclear processes is an intensive numerical calculation requiring discretisation of equation \ref{ntrans} in configuration and velocity space. Invoking spherical symmetry allows multiple interaction types to be retained without restrictive calculation time. With this considered, we have developed a spherical 1D discrete ordinates multi-group neutron transport code, Minotaur. This involves discretisation of the neutron distribution over radius, direction cosine and energy. When cycling over the neutron energy groups, the spatial and angular distribution of neutrons is calculated. Collisions with ions allow transitions between energy and angular groups. Such collision sources are described by the integral double differential cross section term in equation \ref{ntrans}. The collision source is handled numerically using the I*-method \cite{Takahashi1979a,Takahashi1979b} and a von Neumann series solution is converged upon by multiple evaluations of the collision source. Inner iterations within each energy group are used to ensure convergence with the source from within-group transitions\cite{lathrop1965} e.g. from small angle scattering. \\
	
	The birth spectra of fusion neutrons for given plasma conditions were accurately calculated by Appelbe \cite{Appelbe2011,Appelbe2014,Appelbe2016} and are included in this code. Alternatively the relativistic Brysk \cite{Brysk1973,Ballabio1998} spectrum can be used for the DT and DD reactions. This is less accurate but provides an analytic form for the spectra and allows more direct comparison with 3D calculations which also use this form for expedience. Most notably the Brysk model cannot capture the high energy tail present in the more accurate spectra. As most nuclear cross sections decrease with increasing energy above 14 MeV, this high energy tail is not expected to have a significant effect on the spectra at lower energies. \\
	
\begin{figure}[htp]
		\centering
		\includegraphics*[width=0.495\textwidth]{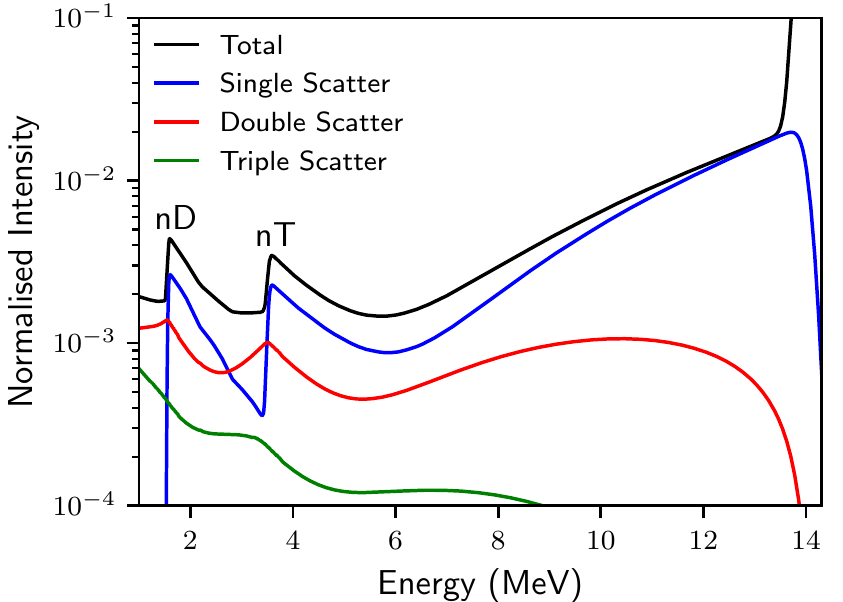}
		\caption{Neutron spectra produced from the down scattering of DT fusion neutrons by deuterons and tritons only. The different contributions from multiple scattering are separated out. A simple isobaric hotspot model\cite{Taylor2013,Taylor2014} was used. The spectra are normalised such that the attenuated DT peak value is unity.}
		\label{fig:increasing_scatterers}
	\end{figure}

	Below the DT peak and down to 10 MeV, the major contribution to the spectrum is from scattering from deuterons and tritons. Scattering from the ablator and D(n,2n) provide minor additions to this region. Multiple scattering from the fuel becomes increasingly important at higher areal densities. Figure \ref{fig:increasing_scatterers} illustrates the spectral contribution of DT multiple scattering at a fuel $\rho R$ of 1 g/cm$^2$. For fuel areal densities relevant to current indirect drive ICF ($\leq$ 1 g/cm$^2$), the 10--12 MeV spectrum is dominated by single scattering from DT. This leads to fitted relations between fuel areal density and DSR \cite{GatuJohnson2016,Frenje2013}. Based on the capsule model used, $\rho R$-DSR relations can be inferred from areal density scans in Minotaur. For two example cases, a point source in a uniform sphere and an isobaric hotspot model \cite{Taylor2014}, linear $\rho$R$_{\mbox{DT}}$-DSR coefficients of 21.0 and 19.4 were found. These are in agreement with others in the literature \cite{GatuJohnson2016,Frenje2013}. To estimate the error in the inverse ray trace method, Minotaur found at 1 g/cm$^2$ the spectrum produced within the single scatter approximation is valid to within < 30\% above 10 MeV causing a 20\% reduction in calculated DSR. These errors increase for larger areal densities and decrease for lower.\\
	
	Spatial information about the production and scattering rates of neutrons was evaluated for a High-Foot simulation. It was found that while 99.9\% of primary neutrons were produced within the 1keV hotspot, only 12\% of DT scattering events occur within this region. Hence the majority of scattering occurs within the cold, dense shell. Possible thermal effects on scattering in the shell will not be considered in our transport models. Inclusion of these effects would require evaluation of the scattering kernel at the relevant ion temperatures \cite{Osborn1958}. This extra dimensionality would increase calculation times.
	
	\section{Backscatter Edge Fitting Models}
	
	This appendix contains the approach and approximations used to obtain the backscatter edge fitting models as well as testing of the fitting model on synthetic data from Minotaur.\\
	
	For elastic scattering with an isotropic centre-of-mass differential cross section and stationary scatterer, the energy spectrum at the backscatter edge can be approximated by:
	\begin{align*}
	I_{bs}(E) \approx& \int_0^{\infty} dE' \frac{d\sigma}{d\Omega_c} g(E') Q_{b}(E') \int d\mu \delta(\mu - \mu^*) \nonumber \\
	=& \int_0^{\infty} dE' \frac{\sigma(E')}{4\pi} g(E') Q_{b}(E') \int d\mu \delta(\mu - \mu^*) \nonumber\\
	=&\int_{E}^{E/\alpha} dE' \frac{\sigma(E')}{4\pi} g(E') Q_{b}(E') \\
	\mu^* = & \frac{1}{2}\left[(A+1)\sqrt{\frac{E}{E'}}-(A-1)\sqrt{\frac{E'}{E}}\right] \nonumber\\
	\alpha = & \left(\frac{A-1}{A+1}\right)^2 \ , \ g(E') = \frac{2}{\left(1-\alpha\right)E'} \nonumber
	\end{align*}
	The backscatter energy reduction factor, $\alpha$, is given above for the case of negligible fluid velocity of the scattering medium . The functional form of the integral can be approximated for a Gaussian birth spectrum, $Q_b(E')$, with a mean $a$ and variance $b^2$. If the birth spectrum is strongly peaked (i.e. $b \ll a$) then the slowing down kernel \cite{takahashi1979}, $g(E')$, can be expanded about $a$ and the elastic scattering cross section taken as linear in energy, i.e. $\sigma \approx \sigma_m E + \sigma_c$. 
	\begin{align*}
	\mbox{Let} \ x &= \frac{E'-a}{b} \\
	\sigma(E') &\approx \sigma_mE'+\sigma_c \rightarrow (\sigma_ma+\sigma_c)+\sigma_mbx\\
	g(E') &\rightarrow \frac{2}{1-\alpha}\left(a+bx\right)^{-1} \approx \frac{2}{(1-\alpha)a}\left(1-\frac{b}{a}x+\frac{b^2}{a^2}x^2\right)\\
	I_{bs}(E) &\propto \int_{\frac{E-a}{b}}^{\frac{E-\alpha a}{\alpha b}} dx  \left[(\sigma_ma+\sigma_c)-\frac{b}{a}\sigma_cx\right] \exp\left[-\frac{x^2}{2}\right]\\
	&+\mathcal{O}\left(\frac{b^2}{a^2}\right)
	\end{align*}
	There is no $\sigma_m$ contribution to the first order in $b/a$ term due to cancellation in the product of the cross section and slowing down kernel. These approximations allow the integral to be performed analytically. To first order in $b/a$, this produces:
	\begin{align*}
	\label{eqnedge}\tag{B0}
	I_{bs}(E) &\propto (\sigma_m a +\sigma_c)\sqrt{\frac{\pi}{2}}\left[\erf\left(\frac{E-\alpha a}{\sqrt{2} \alpha b}\right)+1\right] \nonumber \\
	&+\frac{b}{a}\sigma_ce^{-\frac{(E-\alpha a)^2}{2 (\alpha b)^2}}
	\end{align*}
	Hence information about the mean and variance of the birth spectrum are stored in the form of the backscatter edge. \\
	
\begin{figure}[htp]
		\centering
		\includegraphics*[width=0.495\textwidth]{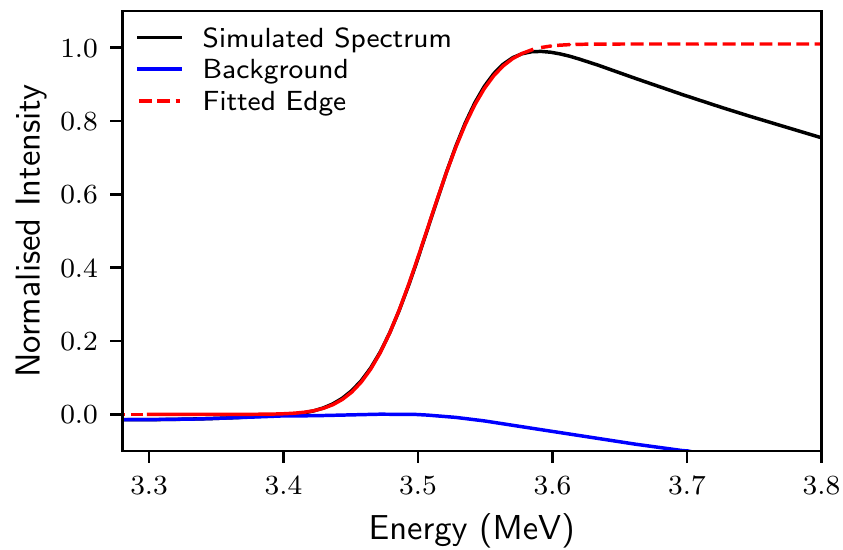}
		\caption{Plot showing a simulated nT edge with backgrounds from scattering from D, multiple scattering, ablator scattering, deuteron break-up and TT primary neutrons. The edge fit is of the form given in equation \ref{eqnedge} plus a constant to fit the background. The solid red line shows the region over which a non-linear least-squares fit was performed. The dashed line shows the value of the fitting function, equation \ref{eqnedge}, beyond this region.}
		\label{fig:nTedgefit}
	\end{figure}

	By using Minotaur, this fit was tested on spectra produced by an isobaric hotspot model \cite{Taylor2014}. Figure \ref{fig:nTedgefit} shows the simulated nT edge and a best fit. The effect of fluid velocity on the broadening on the birth spectra was included. The effects of shell velocity on the edge will be investigated in the following discussion but will be neglected here. A limited energy range around the edge must be used for two reasons. Firstly, the assumption of isotropic centre-of-mass frame scattering will affect the spectral shape above the edge, this will constrain the upper limit of the fitting region. Secondly, using a small energy range will limit the variation in the background sources; over this range they can assumed to be constant, this will constrain the lower limit of the fitting region. The fit was performed from 3.3 MeV up to the peak value of the backscatter edge. Fits extended to higher energy and those which only included a limited section of the edge performed poorly. In these scenarios, the height of the edge was poorly constrained which had a knock on effect on the best fit mean and variance. Hence the fitting region adopted extended from the peak of the edge down to $\geq$ 200 keV below the centre of the edge. From the edge fit the mean and standard deviation of the birth spectrum were calculated to be 14.03 and 0.145 respectively. The primary spectrum moments gave values of 14.05 and 0.147, thus the fit produced good agreement. Since Minotaur does not included the approximations used to obtain model \ref{eqnedge}, the good fit validates the choices made. Slight differences are due to the fact that the spectra detected differ from the birth spectra due to additional interactions on the path to the detector. Differential attenuation causes a shift to higher energies for the primary DT peak due to the cross section decreasing at higher energy. It should be noted that the background signal is relatively constant around the nT edge. Analysis using the single scatter approximation of the 3D ray tracer can precede in the knowledge that background subtraction is feasible in this region. For both scattering and D(n,2n) the n-th order interaction rate $\propto \left(\rho R\right)^n$, therefore at higher areal densities the background may overwhelm the edge, however currently obtainable areal densities are sufficiently low to avoid this. \\

	Often significant fluid velocity, $v_f$, is present in the scattering medium so it is required to extend this spectral backscatter edge fit to moving scatterers. By considering only co-linear collisions between neutrons and the scattering ions we can reformulate to include fluid velocity effects through the energy reduction factor $\alpha$:
	\begin{align}\label{eqnedge1}
		\alpha(v_f \neq 0) &= \left(\frac{A-1}{A+1-2Av_f/v_n}\right)^2 \\
		\mbox{Replace} \ \alpha(v_f &= 0) \rightarrow \alpha(v_f) \ \mbox{in equation \ref{eqnedge}} \nonumber
	\end{align}
	This $\alpha$ is equivalent to equation \ref{eqnalpha} but has been rearranged to remove its $v'_n(E')$ dependence. This will allow a single average shell velocity to be inferred by fitting the backscatter edge.\\
	
	As an extension we will assume that scattering ion velocities sampled by the backscattering neutrons are normally distributed with mean and variance: $\bar{v}_f$ and $\Delta^2_{v}$. This allows us to calculate the expected value of the backscattered spectrum for a range of scattering velocities. For simplicity we will consider only to zeroth order in $b/a$, if the first order $b/a$ terms are retained then this model will account for gradients in the absolute cross section. Expanding $\alpha$ to first order in $v_f/v_n$ we find:
	\begin{align}
	\label{eqnedge2}
		\langle I_{bs}(E) \rangle &= \frac{1}{\sqrt{2\pi\Delta^2_{v}}} \int_{-\infty}^{+\infty} dv_f I_{bs}(E,v_f)  \exp\left[-\frac{\left(v_f-\bar{v}_f\right)^2}{2\Delta^2_{v}}\right] \nonumber \\
		\langle I_{bs}(E) \rangle &\propto 1+\frac{1}{\sqrt{2\pi}}\int_{-\infty}^{+\infty} dy \erf(my+c)\exp[-\frac{y^2}{2}] \\
		c &\equiv \frac{E-\alpha_0 a}{\sqrt{2} \alpha_0 b}+\frac{m}{\Delta_{v}}\bar{v}_f \ , \ m \equiv -\frac{4AE}{\sqrt{2}(A+1)\alpha_0b}\frac{\Delta_{v}}{v_n}\nonumber \\
		\mbox{where} \ &\alpha_0 = \alpha(v_f = 0) \ \mbox{and} \ y = \frac{v_f-\bar{v}_f}{\Delta_{v}}\nonumber
	\end{align}
	Hence the total backscatter shape is simply the sum of edges shifted by variable scattering medium velocity. This will cause an additional smoothing of the resultant edge shape. Thus if model \ref{eqnedge1} were used to fit an edge where there was large variation in the fluid velocity then a larger $b$ value would be obtained. With model \ref{eqnedge2}, now $b$ can be held constant as the birth spectrum variance and $\Delta_v$ can be used to fit the broadening of the edge. It should be noted that the source of this scattering velocity variation, $\Delta_{v}$, is not specified in the model so could be used to treat both finite temperature in the scattering medium and spatial and temporal variation in the fluid velocity. The latter will be studied in this work. \\
	
	Equivalent analysis can be performed on the nD edge although the background may present difficulties. The gradient of the D(n,2n) background is steeper in this region, hence a more sophisticated background model is required. \\
	\section*{References}
	\bibliography{references}
	
\end{document}